\documentclass[%
reprint,
 superscriptaddress,
nofootinbib,
 amsmath,amssymb,
 aps,
floatfix,
]{revtex4-2}

\usepackage{color}
\usepackage{graphicx}
\usepackage{dcolumn}
\usepackage{bm}
\usepackage{booktabs} 
\usepackage{multirow}
\usepackage{makecell}
\usepackage{longtable}

\definecolor{dkgreen}{RGB}{0,100,0}

\begin{document}


\title{Friendship paradoxes in networks}

\author{Eun Lee}
\email{eunlee@pknu.ac.kr}
\affiliation{Department of Scientific Computing, Pukyong National University, Busan, Republic of Korea}

\author{Hang-Hyun Jo}
\email{h2jo@catholic.ac.kr}
\affiliation{Department of Physics, The Catholic University of Korea, Bucheon, Republic of Korea}
\date{\today}

\begin{abstract}
Since Scott L. Feld's seminal paper entitled ``Why your friends have more friends than you do'' [Am.~J.~Sociol.~\textbf{96}, 1464 (1991)], the friendship paradox (FP) has been extensively studied in network science and related fields over the last decades. In this paper we survey the literature on the FP and its generalizations. We first introduce definitions and general results of the FPs defined in networks, and then overview their results on model networks. Empirical results using various social network data as well as applications to real-world problems, such as early detection of outbreaks and network sampling, are also summarized. Finally, we conclude by discussing several directions of possible future works.
\end{abstract}

\maketitle

\tableofcontents

\section{Introduction}\label{sec:intro}

We live in a complex network; individuals interact with each other by trading goods and services, exchanging information, opinions, or rumors, and often spreading viruses. Thus, the interaction structure, depicting who interacts with whom, is of utmost importance to understand various emergent macroscopic patterns of human social behaviors. Network science has provided a powerful framework for describing, explaining, and predicting human social behaviors~\cite{Albert2002Statistical, Boccaletti2006Complex, Borgatti2009Network, Easley2010Networks, Barabasi2016Network, Newman2018Networks, Menczer2020First, Dorogovtsev2025Nature}. A number of social network data have been collected and analyzed to reveal complicated but intriguing patterns of social interactions. Examples of such patterns include heavy-tailed degree distributions~\cite{Barabasi1999Emergence, Jeong2003Complex, Stumpf2012Critical, Broido2019Scalefree, Voitalov2019Scalefree}, degree-degree correlations~\cite{Newman2002Assortative, Newman2003Mixinga}, high clustering and low average path length~\cite{Watts1998Collective, Milgram1967Smallworld}, community structure~\cite{Fortunato2010Community, Schaub2017Many, Javed2018Community, Fortunato202220}, and intensity-topology correlations~\cite{Granovetter1973Strength, Onnela2007Structure, Onnela2007Analysis} among other stylized facts on social networks~\cite{Newman2003Structure, Jo2018Stylized}. In recent years, more comprehensive frameworks for network representation emerged, such as temporal networks~\cite{Holme2012Temporal, Holme2019Temporal, Masuda2020Guide}, multilayer or multiplex networks~\cite{Kivela2014Multilayer, Boccaletti2014Structure, Aleta2019Multilayer, Artime2022Multilayer}, and higher-order networks or hypergraphs~\cite{Lambiotte2019Networks, Battiston2020Networks, Bianconi2021Higherorder, Bick2023What}. All these developments have contributed to the analysis of social network data and to deepen our understanding of them.

One of the interesting phenomena due to such interaction structure of social networks is the friendship paradox (FP). In 1991, Scott L. Feld introduced the FP in his seminal paper~\cite{Feld1991Why}, stating that your friends have on average more friends than you do. Note that Feld did not use the term ``FP" in his original paper\footnote{He did not use the term in his coauthored article published in 2010~\cite{Feld2010Puzzles} either, although it was about various paradoxes and cited his own 1991 paper.}; the term was instead coined by Zuckerman and Jost in 1999~\cite{Zuckerman1999What} and started to appear in the literature only in the beginning of 2010s, e.g., in Christakis and Fowler's 2010 paper~\cite{Christakis2010Social} and Maiya and Berger-Wolf's 2011 paper~\cite{Maiya2011Benefits}. Since then, numerous variants of the FP have been discovered and studied, partly thanks to the access to large-scale digital data on human social behaviors. Extensive developments of the topic in recent years have often led to somewhat confusing terminology for the same concepts, requiring us to sort them out and, more importantly, to summarize the literature in an organized paper.

The underlying mechanism of the FP essentially lies in the heterogeneous degree distribution of the network, where the degree of a node is the number of its neighbors~\cite{Erdos1960Evolution, Freeman1978Centrality}. Just because high-degree nodes have more neighbors than low-degree nodes, they are more often observed by their neighbors than low-degree counterparts, leading to the overrepresentation of high-degree nodes and underrepresentation of low-degree nodes when observing the systems via neighbors. This in turn results in various disparities between groups of high-degree nodes (including hubs) and low-degree nodes. Typically, only a small fraction of nodes having very large degrees attract most of attention from a majority of nodes who have low degrees. 

The most remarkable generalization of the FP might be the generalized FP (GFP)~\cite{Hodas2013Friendship, Eom2014Generalized}. The GFP can be stated such that your friends have on average ``better" attributes than you do\footnote{In this statement, ``better" may be inappropriate in some contexts as it can depend on the definition or perception of being good or bad.}, where the attributes can be derived solely from the network structure (e.g., betweenness centrality~\cite{Freeman1978Centrality}) or they are not necessarily derived from the network topology (e.g., happiness of a social media user~\cite{Bollen2017Happiness}). Simply put, the GFP emerges when degrees and attributes of nodes are correlated with each other~\cite{Eom2014Generalized}. In most cases, higher-degree nodes tend to have better attributes than their lower-degree neighbors who typically have worse attributes than average. Accordingly, the better attributes are more frequently observed, leading to their overrepresentation.

FPs have huge implications in human social behaviors. We tend to compare our own popularity and economic status to those of our friends, family members, and colleagues. Further, how we perceive the entire population beyond our social circles is inevitably biased as we only have limited access to the global information, hence we again rely on our immediate neighbors. These perception biases might affect our personal beliefs and behaviors in our everyday life, such as what to buy and who to vote for. Therefore, studying FPs allows us to better understand microscopic origins or underlying mechanisms behind the macroscopic patterns observed in social network data. In particular, GFPs have shed light on the interplay between network structure and individual nodes' attributes, enabling us to answer the question of how local perceptions of nodal attributes often appear in stark contrast to the global reality. This question has been studied in terms of false consensus~\cite{Mullen1985False, Marks1987Ten, Baer1991Biases}, pluralistic ignorance~\cite{Miller1987Pluralistic, Prentice1993Pluralistic}, and majority illusion~\cite{Lerman2016Majority}, all of which can be better understood in the context of FPs.

To study FPs in networks, a variety of novel notions and characterization methods as well as mathematical, computational models have been developed; concrete network models have been extensively employed to systematically study which features of networks affect the FPs and to what extent. In parallel, various social network data have been collected and analyzed to reveal rich patterns regarding FPs. FPs have also found a number of real-world applications across various domains, spanning from early detection of outbreaks and influence maximization to network sampling and behavioral consequences of perception biases.

In this review, we first introduce definitions and general results in Sec.~\ref{sec:definition} and their results on model networks in Sec.~\ref{sec:model_net}. We then review empirical results in Sec.~\ref{sec:empiric} and summarize applications of FPs in various domains in Sec.~\ref{sec:applications}. Finally, we conclude by discussing the possible future works in Sec.~\ref{sec:discussion}.

\section{Definitions and general results}\label{sec:definition}

\subsection{Individual-level, network-level, and degree-dependent friendship paradoxes}\label{subsec:basic}

Let us consider a network with $N$ individual nodes and $L$ links between them. Although the links may be directed, weighted, higher-ordered, and/or signed depending on the systems of interest, we start with a simple network only with undirected, binary, pairwise, and unsigned links. In addition, self-loops and multiple links between the same pair of nodes are ignored for the sake of simplicity, and we further assume that the network does not contain isolated nodes. The set of nodes in the network is denoted by $V=\{1,\ldots,N\}$, while the set of links is represented by $E=\{(i,j)\}$, where $(i,j)$ or simply $ij$ denotes a link connecting two nodes $i,j\in V$, and $|E|=L$. The degree of a node $i$ for $i\in V$ denotes the number of links adjacent to the node $i$, or equivalently the number of $i$'s neighbors~\cite{Erdos1960Evolution, Freeman1978Centrality}, and it is denoted by $k_i$. 

\begin{figure}[!h]
    \centering
    \includegraphics[width=\columnwidth]{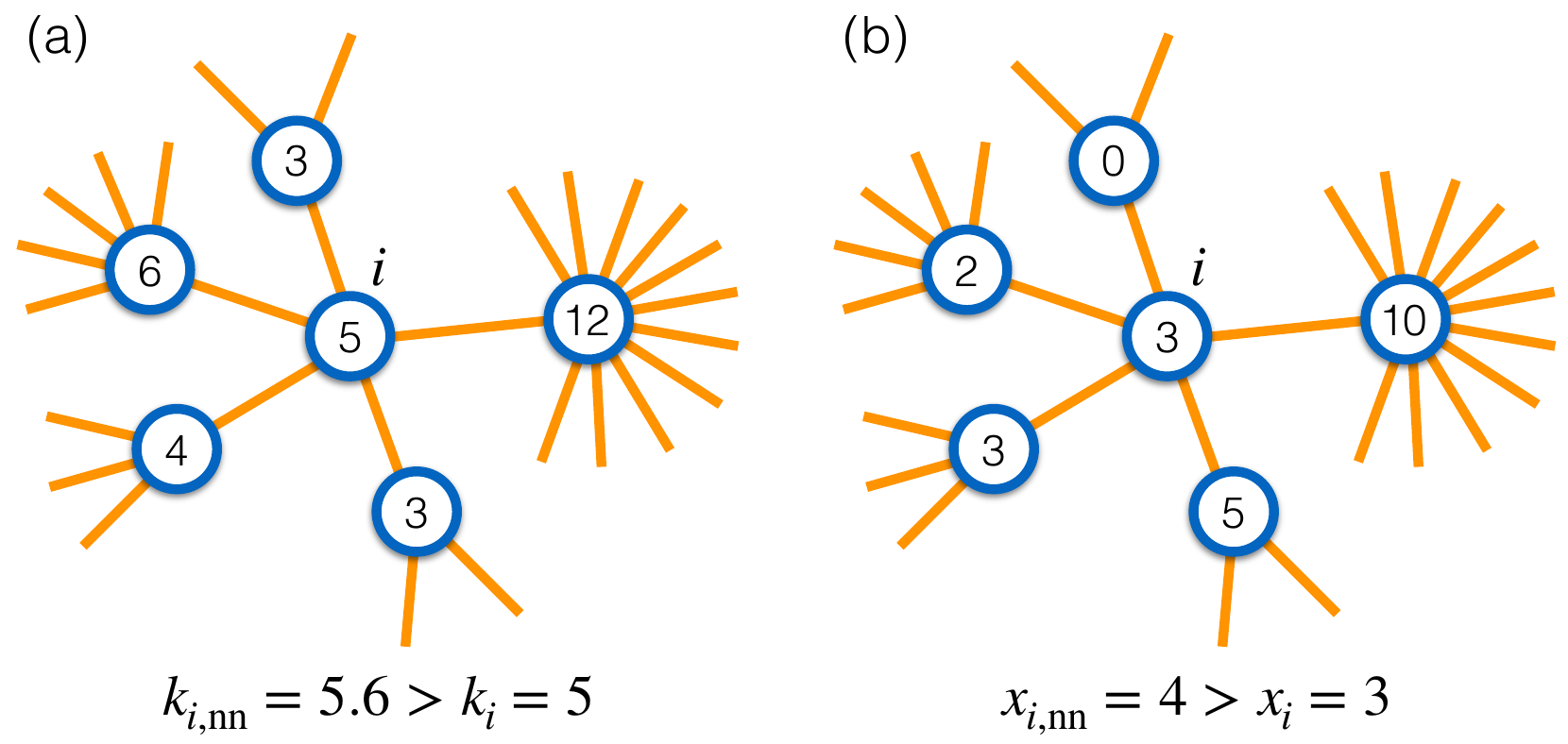}
    \caption{Examples of (a) the friendship paradox (FP) holding for the node $i$ [Eq.~\eqref{eq:FP_ind_define}] and (b) its generalization (GFP) to the nodal attribute [Eq.~\eqref{eq:GFP_ind_define}]. Blue circles and orange lines denote nodes and links. Numbers in circles in panels (a) and (b) represent degrees and attributes of nodes, respectively. See the text for definitions of symbols.}
    \label{fig:schematic}
\end{figure}

To define the friendship paradox (FP) as proposed by Feld~\cite{Feld1991Why}, we define the average degree of $i$'s neighbors as following:
\begin{align}
    k_{i,\rm nn}\equiv \frac{1}{k_i}\sum_{j\in \Lambda_i} k_j,
    \label{eq:k_inn_define}
\end{align}
where $\Lambda_i$ is the set of neighbors of the node $i$, and $|\Lambda_i|=k_i$. The subscript ``nn" means nearest neighbors~\cite{Pastor-Satorras2001Dynamical, Gan2007Data}. Then the FP at the individual level, or ``FP-ind" in short, can be written as\footnote{The FP-ind was also called neighborhood superiority in Ref.~\cite{Momeni2016Qualities}.}
\begin{align}
    k_{i,\rm nn}>k_i.
    \label{eq:FP_ind_define}
\end{align}
That is, the degree of the node $i$ is less than the average degree of its neighbors [see Fig.~\ref{fig:schematic}(a) for an example]. The inequality in Eq.~\eqref{eq:FP_ind_define} does not necessarily hold for all nodes; for example, it may not hold for some nodes having the highest degree in the network. What makes the FP interesting is the fraction of nodes for which the inequality holds, which has been called the network-level paradox holding probability $H$~\cite{Eom2014Generalized, Jo2014Generalized}:
\begin{align}
    H\equiv \frac{1}{N}\sum_i \theta(k_{i,\rm nn}-k_i),
    \label{eq:FP_H_define}
\end{align}
where $\theta(\cdot)$ is the Heaviside step function with the condition $\theta(0)=0$. Let us guess the value of $H$ in some network. Consider a link connecting two nodes $i$ and $j$ who have different degrees such that $k_i<k_j$. This link would increase the chance that $k_{i,\rm nn}>k_i$, but would reduce the chance that $k_{j,\rm nn}>k_j$. Based on such apparent symmetry, one could naively expect that the value of $H$ must be around $1/2$ or less than $1/2$ when there are more links connecting nodes of the same degree. However, most empirical analyses have shown that the value of $H$ is considerably larger than $1/2$ or even very close to $1$ (see Sec.~\ref{sec:empiric}), hence it has been called a paradox. As we will shortly discuss, the observation that $H>1/2$ is not just because of the noise in the data, but due to the profound effects driven by the degree heterogeneity, meaning that some nodes have more neighbors than the others. 

\begin{figure}[!h]
    \centering
    \includegraphics[width=\columnwidth]{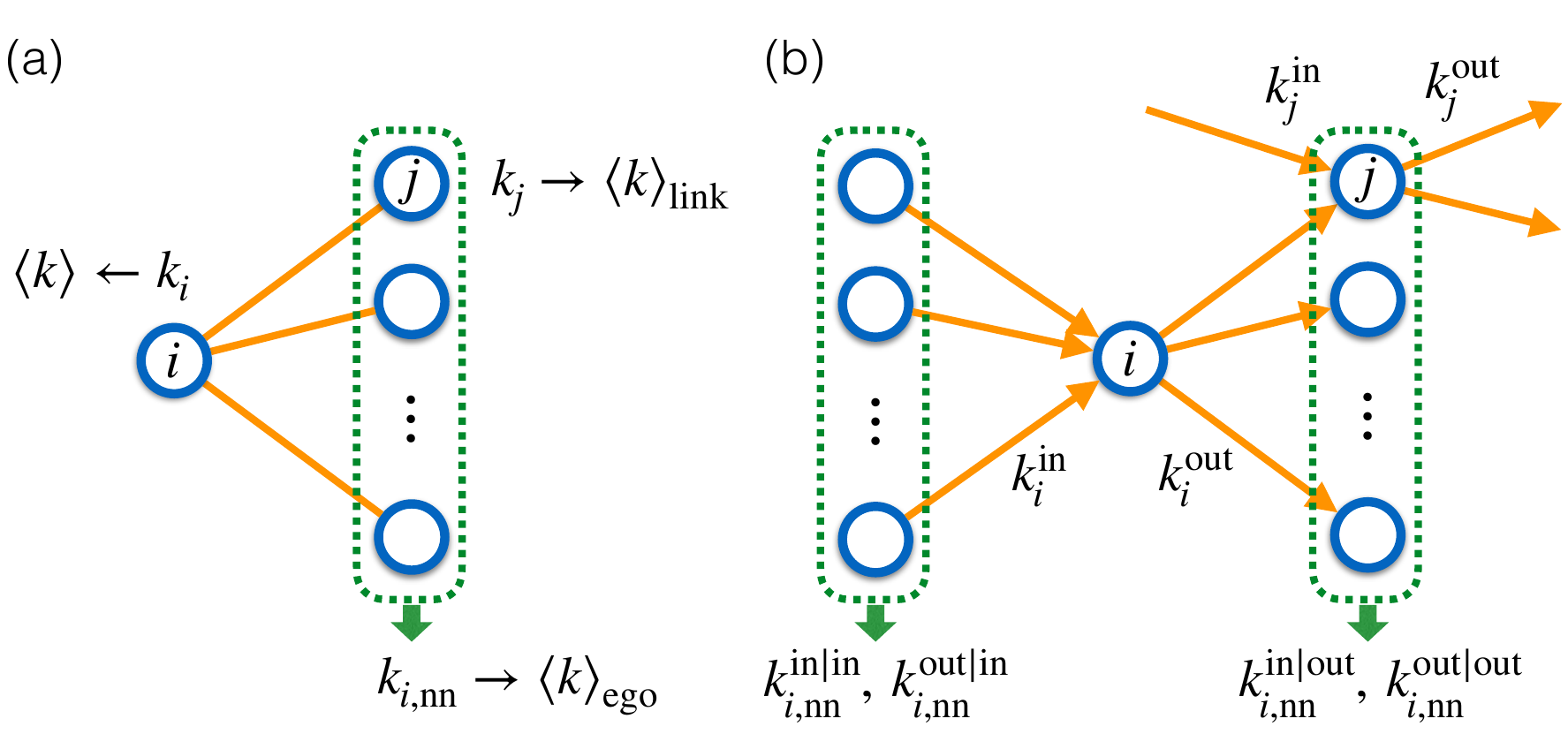}
    \caption{Definitions of degrees and their averages for undirected networks (a) and for directed networks (b). In panel (a), the average degree of the node $i$'s neighbors is denoted by $k_{i,\rm nn}$ in Eq.~\eqref{eq:k_inn_define} and black horizontal arrows indicate averaging over all nodes in the network. In panel (b), the average in-degree/out-degree of in-neighbors/out-neighbors of the node $i$ are denoted by $k_{i,\rm nn}^{\alpha|\beta}$ for $\alpha,\beta\in\{\rm in,out\}$ in Eq.~\eqref{eq:kinn_directed_define}.}
    \label{fig:symbols}
\end{figure}

Feld also defined the network-level FP by comparing the average degree over all nodes to the average degree of neighbors over all nodes~\cite{Feld1991Why}. The average degree and the average degree of neighbors over all nodes are respectively denoted by
\begin{align}
   &\langle k\rangle\equiv \frac{1}{N}\sum_i k_i,\ \\
   &\langle k\rangle_{\rm link} \equiv \frac{\sum_i k_i^2}{\sum_i k_i},
   \label{eq:k_link_define}
\end{align}
where $\langle k\rangle$ equals to the expected degree of a node that is chosen uniformly at random among all nodes. In contrast, $\langle k\rangle_{\rm link}$ is the expected degree of the node reached by a link that is chosen uniformly at random among all links, hence the subscript ``link" [see Fig.~\ref{fig:symbols}(a)]\footnote{The probability of reaching a node with degree $k_i$ following a link that is chosen uniformly at random is given as $p_i\equiv k_i/\sum_j k_j$. Then the expected degree of such a node is obtained as $\sum_i p_i k_i=\sum_i k_i^2/\sum_j k_j=\langle k\rangle_{\rm link}$.}. Here $\langle k\rangle_{\rm link}$ can be mistaken for the expected degree of the random neighbor of the random node, but they are different from each other; the difference will be discussed later. To proceed, let us denote the $n$th moment of the degree as 
\begin{align}
    \langle k^n\rangle \equiv \frac{1}{N}\sum_i k_i^n = \sum_k k^n P(k),
\end{align}
where $P(k)$ is the degree distribution of the network, satisfying the normalization condition, i.e., $\sum_k P(k)=1$. It is worth noting that the probability of reaching a node with degree $k$ following a randomly chosen link is proportional to $kP(k)$ rather than $P(k)$ just because the node with degree $k$ is $k$ times more likely to become somebody's neighbor than those with degree one~\cite{Newman2003Egocentered, Newman2003Why, Newman2007Course}. Such an effect gets more pronounced in networks with heavy-tailed degree distributions such as scale-free networks~\cite{Barabasi1999Emergence, Jeong2003Complex, Stumpf2012Critical, Broido2019Scalefree, Voitalov2019Scalefree}, where some nodes often have hundreds or thousands times more neighbors than the average node does. Then one can rewrite the average degree of neighbors in Eq.~\eqref{eq:k_link_define} as
\begin{align}
   \langle k\rangle_{\rm link} = \frac{\langle k^2\rangle}{\langle k\rangle} = \langle k\rangle + \frac{\sigma_k^2}{\langle k\rangle},
   \label{eq:k_link_result}
\end{align}
where $\sigma_k$ is the standard deviation of degrees. Equation~\eqref{eq:k_link_result} was first derived by Feld~\cite{Feld1991Why} and its derivation has repeatedly appeared in the later publications, e.g., in Refs.~\cite{Newman2007Course, Kramer2016Multistep, Jackson2019Friendship}. Since $\sigma_k^2\geq 0$ by definition, one gets
\begin{align}
     \langle k\rangle_{\rm link}\geq \langle k\rangle.
    \label{eq:k_link_k}
\end{align}
Feld's definition of the network-level FP, or ``FP-net-link" in short, considers only the case that the average degree of neighbors is strictly larger than the average degree, i.e.,
\begin{align}
    \langle k\rangle_{\rm link}>\langle k\rangle.
    \label{eq:FP_net_link_define}
\end{align}
It implies that the FP-net-link holds for any network unless every node has the identical degree as in a regular graph. Throughout this review, we only consider the strict inequality as the condition for the paradox,  based on the sentence ``the mean number of friends of friends is always greater than the mean number of friends of individuals" in the abstract of Feld's original paper~\cite{Feld1991Why}. However, the case of equality was also considered to indicate the FP-net-link, e.g., in Refs.~\cite{Higham2019Centralityfriendship, Higham2026Generalized}. Finally, we note from Eq.~\eqref{eq:k_link_result} that the larger $\sigma_k$ and/or the smaller $\langle k\rangle$ lead to the bigger gap between $\langle k\rangle_{\rm link}$ and $\langle k\rangle$, implying that the FP-net-link gets stronger for sparser networks with more heterogeneous degree distributions\footnote{In this sense, the strength of the FP-net-link can be maximized for a star network where a single central node is connected to all other nodes that are not connected with each other. For a star network with $N$ nodes, as $N$ increases, $\sigma_k$ diverges as $\sqrt{N}$, while $\langle k\rangle$ approaches $2$.}.

We also note the fact that the FP-net-link is a special case of the more general phenomenon called the inspection paradox~\cite{Feld1977Variation, Stein1985Sampling, Ross2003Inspection, Marengo2025Useful}. To state the inspection paradox, let us consider $N$ items, each of which is associated with some positive number. The statistics of such associated numbers can be estimated by sampling items uniformly at random (unbiased samples). However, if the probability of choosing an item is proportional to its associated number, the statistics of associated numbers of sampled items must be biased (biased samples) because items with larger associated numbers will be chosen more likely than those with smaller associated numbers. One can prove that the average of associated numbers of biased samples is no less than that of unbiased samples. Returning to the FP-net-link, $\langle k\rangle$ is the average degree of unbiased samples, while $\langle k\rangle_{\rm link}$ is for the case when the probability of choosing a node is proportional to its degree, i.e., biased samples. Other examples of the inspection paradox include the class size paradox~\cite{Feld1977Variation, Hemenway1982Why} and the waiting time paradox~\cite{Feller2009Introduction,Masuda2020WaitingTime}. The waiting time paradox is important to understanding dynamical processes taking place in temporal networks~\cite{Kivela2012Multiscale, Karsai2018Bursty, Hiraoka2018Correlated, Masuda2020Guide}.

An alternative definition of the network-level FP, or ``FP-net-ego" in short, was proposed by comparing the average of $k_{i,\rm nn}$ over all nodes to the average degree $\langle k\rangle$~\cite{Feld1991Why, Kramer2016Multistep, Jackson2019Friendship, Cantwell2021Friendship}. The FP-net-ego is said to hold when
\begin{align}
    \langle k\rangle_{\rm ego} \equiv \frac{1}{N}\sum_i k_{i,\rm nn}> \langle k\rangle,
    \label{eq:FP_net_ego_define}
\end{align}
where $\langle k\rangle_{\rm ego}$ is the expected degree of a randomly chosen neighbor of a randomly chosen node, hence the subscript ``ego" [see Fig.~\ref{fig:symbols}(a)]\footnote{The probability of choosing a node $j$ randomly among the neighbors of a randomly chosen node $i$ is given as $p_{j|i}\equiv 1/(Nk_i)$ with $k_i$ denoting the $i$'s degree. Then the expected degree of such a node is obtained as $\sum'_{ij} p_{j|i} k_j=\sum'_{ij} k_j/(Nk_i)=\langle k\rangle_{\rm ego}$, where the summation $\sum'$ is taken over all links. Note that the FP-net-ego was considered to hold even in the case of the equality, i.e., $\langle k\rangle_{\rm ego}=\langle k\rangle$, e.g., in Ref.~\cite{Hazra2026Generalized}.}. Kramer et al.~\cite{Kramer2016Multistep}, Jackson~\cite{Jackson2019Friendship}, and later Cantwell et al.~\cite{Cantwell2021Friendship} independently proved that
\begin{align}
    \langle k\rangle_{\rm ego}\geq \langle k\rangle,
    \label{eq:FP-net-ego_proof}
\end{align}
where the equality holds when the network is locally regular, meaning that each link in the network connects a pair of nodes having the same degree\footnote{For a connected network, the equality holds for a regular graph. When the network consists of multiple connected components, the equality holds when each connected component is regular, that is, all nodes in the same component have the same degree.}. To study properties of $\langle k\rangle_{\rm ego}$ in more detail, the weighted degree of the node $i$ is defined as the sum of the inverse degrees of $i$'s neighbors~\cite{Cantwell2021Friendship, Evtushenko2023Nodebased, Bhattacharya2026Triangle}:
\begin{align}
    \kappa_i \equiv \sum_{j\in \Lambda_i}\frac{1}{k_j}.
    \label{eq:weighted_degree}
\end{align}
Also, the Pearson correlation coefficient (PCC) between $\{k_i\}$ and $\{\kappa_i\}$ is written as
\begin{align}
    \rho_{k\kappa}\equiv \frac{1}{N}\sum_i \frac{(k_i-\langle k\rangle)(\kappa_i-\langle \kappa\rangle)}{\sigma_k\sigma_\kappa},
    \label{eq:rho_k_kappa}
\end{align}
where $\langle\kappa\rangle$ and $\sigma_\kappa$ are the mean and standard deviation of $\{\kappa_i\}$, respectively. Cantwell et al.~\cite{Cantwell2021Friendship} proved that
\begin{align}
    &\langle\kappa\rangle=1,\ \\
    &\rho_{k\kappa}\geq 0.
    \label{eq:rho_kkappa_result}
\end{align}
Then, Eq.~\eqref{eq:FP-net-ego_proof} can be understood by the relation that
\begin{align}
    \langle k\rangle_{\rm ego}=\langle k\rangle+\rho_{k\kappa}\sigma_k\sigma_\kappa,
    \label{eq:k_ego_result}
\end{align}
which was not explicitly shown in Ref.~\cite{Cantwell2021Friendship} but can be easily derived. The larger $\sigma_k$, $\sigma_\kappa$, and/or $\rho_{k\kappa}$ lead to the bigger gap between $\langle k\rangle_{\rm ego}$ and $\langle k\rangle$, implying stronger effects of the FP-net-ego. 

In conclusion, both definitions of the network-level FP, i.e., FP-net-link in Eq.~\eqref{eq:FP_net_link_define} and FP-net-ego in Eq.~\eqref{eq:FP_net_ego_define}, are legitimate formulations of the statement that ``your friends have on average more friends than you do''. Also, both turn out to be generic consequences rooted in the degree heterogeneity of the network. Yet $\langle k\rangle_{\rm link}$ and $\langle k\rangle_{\rm ego}$ are different from each other\footnote{The difference between $\langle k\rangle_{\rm ego}$ and $\langle k\rangle_{\rm link}$ was pointed out in Ref.~\cite{Cao2016Friendship}. $\langle k\rangle_{\rm ego}$ and $\langle k\rangle_{\rm link}$ were alternatively called ego-based and alter-based means in Ref.~\cite{Kumar2024Friendship}, node-averaged and edge-averaged means in Ref.~\cite{Lee2026Two}, and local and global averages in Ref.~\cite{Higham2026Generalized}. In addition, the FP-net-ego and FP-net-link were respectively called singular and list versions of the FP in Ref.~\cite{Evtushenko2024Comprehensive}.}. As Kumar et al.~\cite{Kumar2024Friendship} pointed out, $\langle k\rangle_{\rm link}$ depends only on the degree distribution irrespective of other network properties, whereas $\langle k\rangle_{\rm ego}$ can vary with the network configuration while keeping the same degree distribution. Suppose we have a network with a fixed degree sequence $K\equiv \{k_1,\ldots,k_N\}$, i.e., with a fixed value of $\langle k\rangle_{\rm link}$. One can generate an ensemble of networks by rewiring links while keeping $K$ intact; these networks may have different weighted degree sequences $\{\kappa_1,\ldots,\kappa_N\}$, leading to different values of $\rho_{k\kappa}$ and $\sigma_\kappa$, hence different values of $\langle k\rangle_{\rm ego}$. The network-level paradox holding probability $H$ also depends on the network configuration in a similar manner\footnote{Lattanzi and Singer~\cite{Lattanzi2015Power} conceived a kind of networks for which the FP-net-link holds but $H=0$. When a network is given as a set of disjoint cliques of different sizes, each node has neighbors with exactly the same degree of the node, i.e., $H=0$, whereas the variance of degrees of the entire network is positive, implying the FP-net-link. Yet most empirical networks are hardly described by disjoint cliques, or only the largest connected components are considered, naturally avoiding such an issue.}. 

To be precise, one can readily compute the difference between $\langle k\rangle_{\rm link}$ and $\langle k\rangle_{\rm ego}$ using Eqs.~\eqref{eq:k_link_result} and~\eqref{eq:k_ego_result} as
\begin{align}
    \langle k\rangle_{\rm ego} - \langle k\rangle_{\rm link} = \sigma_k\sigma_\kappa\left(\rho_{k\kappa} - \frac{\sigma_k/\langle k\rangle}{ \sigma_\kappa}\right).
    \label{eq:diff_ego_link_review}
\end{align}
However, this result is not so intuitive to understand. Recently, Kumar et al.~\cite{Kumar2024Friendship} derived such a difference as follows:
\begin{align}
    \langle k\rangle_{\rm ego} - \langle k\rangle_{\rm link} = \rho_{kk^{-1}}
    \frac{\sqrt{(\langle k\rangle\langle k^3\rangle-\langle k^2\rangle^2)(\langle k\rangle\langle k^{-1}\rangle-1)}}{\langle k\rangle},
    \label{eq:diff_ego_link_Kumar}
\end{align}
where $\rho_{kk^{-1}}$ denotes the PCC between the degree of one end node and the inverse degree of the other end node of the link, and it is called inversity. Which of $\langle k\rangle_{\rm ego}$ and $\langle k\rangle_{\rm link}$ is bigger depends on the sign of the inversity. More recently, Lee~\cite{Lee2026Two} derived the same difference but in terms of the PCC between the degrees of nodes and the average degrees of their neighbors:
\begin{align}
    \langle k\rangle_{\rm ego} - \langle k\rangle_{\rm link} = - \frac{\rho_{kk_{\rm nn}}\sigma_k \sigma_{k_{\rm nn}}}{\langle k\rangle},
    \label{eq:diff_ego_link_Lee}
\end{align}
where $\rho_{kk_{\rm nn}}$ is the PCC between $\{k_i\}$ and $\{k_{i,{\rm nn}}\}$, and $\sigma_{k_{\rm nn}}$ denotes the standard deviation of $\{k_{i,{\rm nn}}\}$. These three relations in Eqs.~\eqref{eq:diff_ego_link_review}--\eqref{eq:diff_ego_link_Lee} are equivalent to each other but only differ as to which PCC is used to quantify the degree-degree correlation. 

In addition to the individual-level and network-level FPs, one can study the FP at the intermediate or mesoscopic level. Eom and Jo~\cite{Eom2014Generalized, Jo2014Generalized} introduced the paradox holding probability for a group of nodes with the same degree, namely, the degree-dependent FP. Let $V_k$ denote the set of nodes with the degree $k$ and the paradox holding probability for this set is denoted by $h(k)$:
\begin{align}
    h(k) \equiv \frac{1}{|V_k|}\sum_{i\in V_k} \theta(k_{i,\rm nn}-k).
    \label{eq:FP_hk_define}
\end{align}
The degree-dependent FP provides the stratified understanding of the FP, which is particularly useful for networks with heavy-tailed degree distributions. For a network with the degree distribution $P(k)$, one can relate $h(k)$ to $H$ such that
\begin{align}
    H=\sum_k h(k) P(k).
\end{align}
Alternatively, Ugander et al.~\cite{Ugander2011Anatomy} compared the average degree of neighbors of nodes with degree $k$ to the degree $k$. Such an average degree of neighbors is denoted by
\begin{align}
    \langle k\rangle_{\rm nn}(k)\equiv \frac{1}{|V_k|}\sum_{i\in V_k}k_{i,\rm nn}= \frac{1}{|V_k|k}\sum_{i\in V_k}\sum_{j\in\Lambda_i}k_j=\sum_{k'}k'P(k'|k),
    \label{eq:k_nn_k}
\end{align}
where $P(k'|k)$ is a conditional probability distribution of $k'$ being the degree of a neighbor of the node with degree $k$~\cite{Maslov2004Detection}. If $\langle k\rangle_{\rm nn}(k)>k$ for some $k$, it means that the FP holds within the group of nodes with degree $k$. The average degree of neighbors in Eq.~\eqref{eq:k_nn_k} was originally introduced to characterize the degree assortativity of the network~\cite{Pastor-Satorras2001Dynamical, Onnela2007Analysis, Smith2008Linkspace}; the increasing (decreasing) behavior of $\langle k\rangle_{\rm nn}(k)$ as a function of $k$ indicates the tendency of connecting nodes with similar (different) degrees, i.e., homophily (heterophily).

Finally, we brief two extensions of the FP, namely, (i) a multiple-step or multistep FP and (ii) the FP for common neighbors. The multistep FP states that your friends' friends' $\ldots$ friends have on average more friends than you do~\cite{Kramer2016Multistep, Berenhaut2018Degreewise, Berenhaut2024Disparitypersistence, Hazra2026Multilevel}, which can be formulated using random walks~\cite{Pearson1905Problem, Noh2004Random, Masuda2017Random}. A random walk of length $l$ starting from a node $v_0$ is denoted by $(v_0,\ldots,v_l)$. Here $v_{i+1}$ is a random neighbor of the node $v_i$ for $i=0,\ldots,l-1$. When $v_0$ is chosen uniformly at random among all nodes in the network, the expected degree of $v_0$, i.e., $\langle k\rangle$, is shown to be no higher than the expected degree of $v_l$ for every length $l$, hence the multistep extension of the FP-net-ego. On the other hand, the multistep version of the FP-net-link was also considered; one first prepares the set of all paths of length $l$ and choose one path uniformly at random, whose last node is denoted by $v_l$. The expected degree of the node $v_l$ for the odd $l$ is no smaller than $\langle k\rangle$, whereas this is not always the case with even values of $l$~\cite{Kramer2016Multistep, Berenhaut2024Disparitypersistence}. More recently, Xu et al.~\cite{Xu2025Modeling} studied the multistep FP in evolving social networks. It is remarkable that the multistep FP reveals a deep connection of the FP to the random walk problem, implying that any network methodology based on random walks must inherit sampling bias due to the FP (e.g., see Ref.~\cite{Krapivsky2024Magic}).

Next, McGail and Feld~\cite{McGail2026Why} extended the FP to the case with common neighbors; they considered a common neighbor to $l+1$ randomly chosen nodes, equivalently, a node commonly reached by $l+1$ randomly chosen links ($l\geq 0$). The expected degree of such a node, denoted by $\langle k\rangle_{l+1}$, is derived using the mean and variance of the degree of a common neighbor to $l$ random nodes, respectively denoted by $\langle k\rangle_{l}$ and $\sigma^2_{k,l}$, as follows:
\begin{align}
    \langle k\rangle_{l+1}= \langle k\rangle_l+\frac{\sigma^2_{k,l}}{\langle k\rangle_l-l}.
\label{eq:mcgail_eq}
\end{align}
The case with $l=0$ reduces to Eq.~\eqref{eq:k_link_result} as $\langle k\rangle_0=\langle k\rangle$, $\sigma^2_{k,0}=\sigma^2_k$, and $\langle k\rangle_1=\langle k\rangle_{\rm link}$ by definition. In general, one finds that $\langle k\rangle_{l+1}\geq \langle k\rangle_{l}$, implying that a common neighbor to more people tends to have a higher degree. This tendency gets more pronounced for networks with larger $\sigma^2_{k,l}$ and smaller $\langle k\rangle_l$.

\subsection{Generalized friendship paradoxes}\label{subsec:gfp}

Individuals embedded in a network often compare their own attributes with those of their neighbors. Such attributes can be either topological or nontopological. Topological attributes of nodes are those derived only by the topological structure of the network. Examples of topological attributes include various network-centrality measures such as the degree, closeness, betweenness, eigenvector, Katz, PageRank, and walk-count centralities~\cite{Freeman1978Centrality, Feld1991Why, Grund2014Why, Higham2019Centralityfriendship, Kong2019How, Higham2026Generalized, Hazra2026Generalized, Bhattacharya2026Generalized}, as well as the number of triangles attached to the node~\cite{Bhattacharya2026Triangle}. Nontopological attributes of nodes are not necessarily derived from the network structure; examples include the activity and the number of received ``likes" of an online social network user, the number of publications and h-index of a researcher, subjective well-being such as happiness and sentiment, and susceptibility to online influence~\cite{Hodas2013Friendship, Eom2014Generalized, Scissors2016Whats, Benevenuto2016Hindex, Bollen2017Happiness, Zhou2020Sentiment, Arugute2023Network, Luceri2025Susceptibility}. Nontopological attributes can be either numeric or categorical~\cite{Park2007Distribution, Meng2016Interplay, Roy2025Homophilic}, though we focus on numeric attributes in this review unless otherwise stated. Then, the generalized friendship paradox (GFP) can be stated such that your friends have on average better attributes than you do~\cite{Eom2014Generalized}. 

Precisely, let $x_i$ denote the attribute associated with or assigned to the node $i$. The attribute distribution is denoted by $P(x)$. The GFP at the individual level, or ``GFP-ind" in short, can be written as [see Fig.~\ref{fig:schematic}(b) for an example]
\begin{align}
    x_{i,\rm nn}\equiv \frac{1}{k_i}\sum_{j\in \Lambda_i} x_j>x_i,
    \label{eq:GFP_ind_define}
\end{align}
where $x_{i,\rm nn}$ is the average attribute of $i$'s neighbors. Similarly to $H$ in Eq.~\eqref{eq:FP_H_define}, one can compute the fraction of nodes for which the inequality in Eq.~\eqref{eq:GFP_ind_define} holds~\cite{Eom2014Generalized}:
\begin{align}
    H_{\rm at}\equiv \frac{1}{N}\sum_i \theta(x_{i,\rm nn}-x_i).
    \label{eq:GFP_H_define}
\end{align}

Next, we introduce two versions of the GFP at the network level, namely, GFP-net-link and GFP-net-ego, for which the following quantities are defined:
\begin{align}
   &\langle x\rangle\equiv \frac{1}{N}\sum_i x_i,\ \\
   &\langle x\rangle_{\rm link} \equiv \frac{\sum_i k_i x_i}{\sum_i k_i},\  \\
   &\langle x\rangle_{\rm ego}\equiv \frac{1}{N}\sum_i x_{i,\rm nn},
   \label{eq:x_link_x_ego}
\end{align}
where $\langle x\rangle$ is equal to the expected attribute of a randomly chosen node, $\langle x\rangle_{\rm link}$ is the expected attribute of the node reached by a randomly chosen link, and $\langle x\rangle_{\rm ego}$ is the expected attribute of a random neighbor of a random node. The GFP-net-link is written as follows~\cite{Eom2014Generalized}:
\begin{align}
    \langle x\rangle_{\rm link}>\langle x\rangle,
    \label{eq:GFP_net_link_define}
\end{align}
while the GFP-net-ego is expressed as the following inequality~\cite{Cantwell2021Friendship, Evtushenko2024Comprehensive}:
\begin{align}
    \langle x\rangle_{\rm ego}> \langle x\rangle.
    \label{eq:GFP_net_ego_define}
\end{align}

\begin{table*}[!ht]
\caption{Summary of network-level quantities.}
    \begin{tabular}{l|ll|ll}
    \hline
    FP & $\displaystyle\langle k\rangle_{\rm link}=\langle k\rangle+\frac{\sigma^2_k}{\langle k\rangle}$ & Eq.~\eqref{eq:k_link_result} & $\displaystyle\langle k\rangle_{\rm ego}=\langle k\rangle+\rho_{k\kappa}\sigma_k\sigma_\kappa$ & Eq.~\eqref{eq:k_ego_result}
    \\ \hline
    GFP & $\displaystyle\langle x\rangle_{\rm link}=\langle x\rangle+\frac{\rho_{kx}\sigma_k\sigma_x}{\langle k\rangle}$ & Eq.~\eqref{eq:GFP-net-link} & $\displaystyle\langle x\rangle_{\rm ego}=\langle x\rangle+\rho_{\kappa x}\sigma_\kappa\sigma_x$ & Eq.~\eqref{eq:GFP-net-ego} \\
    \hline
    \end{tabular}
    \label{table:net_quantity}
\end{table*}

To understand the origin of GFPs at the network level, the PCC between degrees and attributes of nodes is introduced as
\begin{align}
    \rho_{kx}\equiv \frac{1}{N}\sum_i \frac{(k_i-\langle k\rangle)(x_i-\langle x\rangle)}{\sigma_k\sigma_x},
    \label{eq:rho_kx}
\end{align}
where $\sigma_x$ is the standard deviation of $\{x_i\}$. The PCC between weighted degrees [$\kappa_i$ in Eq.~\eqref{eq:weighted_degree}] and attributes of nodes is also defined as
\begin{align}
    \rho_{\kappa x}\equiv \frac{1}{N}\sum_i \frac{(\kappa_i-\langle\kappa\rangle)(x_i-\langle x\rangle)}{\sigma_\kappa \sigma_x}.
    \label{eq:rho_kappax}
\end{align}
Then, one gets~\cite{Eom2014Generalized}
\begin{align}
    \langle x\rangle_{\rm link} = \langle x\rangle + \frac{\rho_{kx}\sigma_k\sigma_x}{\langle k\rangle},
    \label{eq:GFP-net-link}
\end{align}
and~\cite{Cantwell2021Friendship}
\begin{align}
    \langle x\rangle_{\rm ego}=\langle x\rangle+\rho_{\kappa x}\sigma_\kappa\sigma_x.
    \label{eq:GFP-net-ego}
\end{align}
We immediately make the following two observations: (i) If the network is regular, i.e., all nodes have the same degree ($\sigma_k=\sigma_\kappa=0$), no network-level GFPs hold because attributes in the network are equally sampled regardless of whether the sampling is uniform, link-based, or ego-based. (ii) If all attributes are identical to $x$ ($\sigma_x=0$), no network-level GFPs hold either, because the expected attribute of sampled nodes is the same as $x$ regardless of the sampling method. Therefore, for any irregular, attributed network with $\sigma_x,\sigma_k>0$, the GFP-net-link holds for the network if and only if $\rho_{kx}>0$. Similarly, for any irregular, attributed network with $\sigma_x,\sigma_\kappa>0$, the GFP-net-ego holds for the network if and only if $\rho_{\kappa x}>0$. In other words, if the attributes of nodes are positively correlated with their degrees (weighted degrees), the GFP-net-link (GFP-net-ego) holds for the network in the same way as the network-level FPs hold for the network. In contrast, if the attributes of nodes are negatively correlated with their degrees or weighted degrees, one finds a tendency opposite to the GFP, namely, $\langle x\rangle_{\rm link} < \langle x\rangle$ or $\langle x\rangle_{\rm ego} < \langle x\rangle$, which was called an anti-GFP~\cite{Eom2014Generalized}. The anti-GFP states that your friends have on average worse attributes than you do. Network-level quantities for both FPs and GFPs are summarized in Table~\ref{table:net_quantity}. 

We now consider the mesoscopic description of the GFP in terms of a degree- and attribute-dependent paradox holding probability $h(k,x)$~\cite{Eom2014Generalized, Jo2014Generalized}, which is defined as
\begin{align}
    h(k,x) \equiv \frac{1}{|V_{k,x}|}\sum_{i\in V_{k,x}} \theta(x_{i,\rm nn}-x),
    \label{eq:GFP_hkx_define}
\end{align}
where $V_{k,x}$ denotes the set of nodes with degree $k$ and attribute $x$. If the probability of finding a node with degree $k$ and attribute $x$ is $P(k,x)$, one can relate $h(k,x)$ to $H_{\rm at}$ in Eq.~\eqref{eq:GFP_H_define} such that
\begin{align}
    H_{\rm at}=\sum_{k,x} h(k,x) P(k,x).
\end{align}
Then, degree-dependent and attribute-dependent paradox holding probabilities are respectively defined by
\begin{align}
    &h_{\rm at}(k) \equiv \frac{1}{|V_k|}\sum_{i\in V_k} \theta(x_{i,\rm nn}-x_i),\\
    &h(x) \equiv \frac{1}{|V_x|}\sum_{i\in V_x} \theta(x_{i,\rm nn}-x),
    \label{eq:GFP_hk_hx_define}
\end{align}
where $V_x$ denotes the set of nodes with attribute $x$.

For the attribute-dependent GFP, one can measure the average attribute of neighbors of nodes having the same attribute $x$:
\begin{align}
    \langle x\rangle_{\rm nn}(x)\equiv \frac{1}{|V_x|}\sum_{i\in V_x}x_{i,\rm nn}= \frac{1}{|V_x|}\sum_{i\in V_x}\frac{1}{k_i}\sum_{j\in\Lambda_i}x_j.
    \label{eq:x_nn_x}
\end{align}
If $\langle x\rangle_{\rm nn}(x)>x$ for some $x$, it means that the GFP holds within the group of nodes with attribute $x$. Similarly, the average attribute of neighbors of nodes with degree $k$ is defined as
\begin{align}
    &\langle x\rangle_{\rm nn}(k)\equiv \frac{1}{|V_k|}\sum_{i\in V_k}x_{i,\rm nn}= \frac{1}{|V_k|k}\sum_{i\in V_k}\sum_{j\in\Lambda_i}x_j \notag\\
    &=\sum_{x'}x'P(x'|k),
    \label{eq:x_nn_k}
\end{align}
where $P(x'|k)$ is a conditional probability distribution of $x'$ being the attribute of a neighbor of the node with degree $k$. For comparison, the average attribute of nodes with degree $k$ is defined as
\begin{align}
    \langle x\rangle(k)\equiv \frac{1}{|V_k|}\sum_{i\in V_k} x_i.
    \label{eq:avg_x_k}
\end{align}
If $\langle x\rangle_{\rm nn}(k)>\langle x\rangle(k)$ for some $k$, it means that the GFP holds within the group of nodes with degree $k$~\cite{Kong2019How}. Note that by setting $x_i=k_i$ for all $i$, each GFP reduces to its corresponding FP.

Finally, the mentioned network-level FPs and GFPs can be expressed and studied using linear algebra. An $N\times N$ adjacency matrix $A$ for a network has elements of $a_{ij}$, which has the value of $1$ if the link $ij$ exists, and $0$ otherwise. Then, various centrality measures of nodes, i.e., topological nodal attributes, can be derived from $A$~\cite{Estrada2010Network}. For example, the degree centrality vector is obtained as ${\bf d} = A{\bf 1}$, where ${\bf 1}$ is a column vector of $1$s. Using these notations, e.g., the FP-net-link reads~\cite{Higham2019Centralityfriendship}
\begin{align}
    \frac{{\bf d}^\top{\bf d}}{{\bf 1}^\top{\bf d}} - \frac{{\bf 1}^\top{\bf d}}{N}> 0,
\end{align}
where $^{\top}$ means the transposition. Similarly, the GFP-net-link can be written as
\begin{align}
    \frac{{\bf d}^\top{\bf x}}{{\bf 1}^\top{\bf d}} - \frac{{\bf 1}^\top{\bf x}}{N}> 0,
\end{align}
where ${\bf x}$ is a column vector of nodal attributes $x_i$. For example, ${\bf x}$ can denote the eigenvector centralities of nodes in a connected network, which is given by  $A{\bf x}=\lambda_1 {\bf x}$, with $\lambda_1$ denoting the largest eigenvalue of $A$. Interested readers may refer to Refs.~\cite{Higham2019Centralityfriendship, Hazra2026Generalized, Higham2026Generalized} for more details of this approach.

\subsection{Friendship index and neighborhood gaps}\label{subsec:gap}

The original FP and their extensions have been given in the form of inequality. However, an inequality does not explicitly reveal the difference or gap between its two sides, which would help us understand the phenomenon more precisely. To quantify the strength or intensity of the individual-level FP, Pal et al.~\cite{Pal2019Quantifying, Pal2019Study} introduced the friendship index for a node, which is defined as the ratio of the average degree of neighbors of the node $i$ to the $i$'s own degree:
\begin{align}
    I_i\equiv \frac{k_{i,\rm nn}}{k_i}=\frac{1}{k_i^2}\sum_{j\in \Lambda_i} k_j.
    \label{eq:FI_define}
\end{align}
Note that Hodas et al.~\cite{Hodas2013Friendship} proposed the similar definitions for the directed network, and that Jackson~\cite{Jackson2019Friendship} used the same quantity for the empirical analysis. Instead of the ratio of $k_{i,\rm nn}$ and $k_i$, Cantwell et al.~\cite{Cantwell2021Friendship} introduced their difference as
\begin{align}
    \Delta k_i\equiv k_{i,\rm nn}-k_i.
    \label{eq:deltak_i}
\end{align}
The individual-level FP holds for the node $i$ if and only if $\Delta k_i>0$. The quantity $\Delta k_i$ has been called a gap~\cite{Evtushenko2024Comprehensive} or a friendship bias~\cite{Hazra2026Friendship, Hazra2026Friendshipa, Bhattacharya2026Triangle, Hollander2026Large}. Here, we suggest another, hopefully clearer name for it, i.e., a neighborhood gap. Since the neighborhood gap and the friendship index are exchangeable by the relation that
\begin{align}
    I_i=1+\frac{\Delta k_i}{k_i},   
\end{align}
we will focus on the neighborhood gap hereafter. For a given network, one can calculate the distribution of neighborhood gaps, $P(\Delta k)$, to study the FP in more detail. Note that the probability of $\Delta k>0$ or $\Pr[\Delta k>0]$ is equivalent to $H$ in Eq.~\eqref{eq:FP_H_define}. From Eq.~\eqref{eq:FP-net-ego_proof}, it is obvious that the average neighborhood gap is nonnegative, namely,
\begin{align}
    \langle \Delta k\rangle_{\rm ego} \equiv \frac{1}{N}\sum_i \Delta k_i = \langle k\rangle_{\rm ego} - \langle k\rangle\geq 0,
    \label{eq:Delta_ego}
\end{align}
where the equality holds only when the network is locally regular. It also means that the FP-net-ego holds for any network that is not locally regular. Another network-level neighborhood gap using $\langle k\rangle_{\rm link}$ was proposed as~\cite{Fotouhi2015Generalized, Evtushenko2024Comprehensive}
\begin{align}
    \langle \Delta k\rangle_{\rm link}\equiv \langle k\rangle_{\rm link} - \langle k\rangle.
    \label{eq:FP_gap_link_define}
\end{align}
By Eq.~\eqref{eq:k_link_k}, it is straightforward to find that $\langle\Delta k\rangle_{\rm link}\ge 0$, and the value of this network-level neighborhood gap tells how strong the FP is at the network level. Recently, Roga~\cite{Roga2026Friendship} defined the network-level neighborhood gap in terms of the expected neighborhood gap of a node sampled with probability proportional to its degree, and showed that the expected neighborhood gap vanishes, concluding that the FP disappears under such sampling.

All the above definitions for the FP can be naturally extended to the GFP~\cite{Cantwell2021Friendship, Evtushenko2024Comprehensive}. Cantwell et al.~\cite{Cantwell2021Friendship} introduced the neighborhood gap for a node $i$ but using its own and its neighbors' attributes:
\begin{align}
    \Delta x_i \equiv x_{i,\rm nn}-x_i.
    \label{eq:deltax_i_define}
\end{align} 
The individual-level GFP holds for the node $i$ if and only if $\Delta x_i>0$. There are two versions of the network-level neighborhood gap for attributes. The first is defined as the average of $\Delta x_i$ as follows:
\begin{align}
    \langle\Delta x\rangle_{\rm ego} \equiv \frac{1}{N}\sum_i \Delta x_i = \langle x\rangle_{\rm ego}-\langle x\rangle =\rho_{\kappa x}\sigma_\kappa \sigma_x.
    \label{eq:GFP_gap_ego_define}
\end{align}
We remind that $\kappa_i$ is the sum of inverse degrees of $i$'s neighbors in Eq.~\eqref{eq:weighted_degree}, called the weighted degree. Thus, the value of $\langle\Delta x\rangle_{\rm ego}$ is positive if and only if the GFP-net-ego holds for the network. The second version of the network-level neighborhood gap is given by
\begin{align}
    \langle\Delta x\rangle_{\rm link}\equiv \langle x\rangle_{\rm link} - \langle x\rangle= \frac{\rho_{kx}\sigma_k\sigma_x}{\langle k\rangle}.
    \label{eq:GFP_gap_link_define}
\end{align}
The value of $\langle\Delta x\rangle_{\rm link}$ is positive if and only if the GFP-net-link holds for the network. $\langle\Delta x\rangle_{\rm link}$ was also called the network-level quality paradox in Ref.~\cite{Fotouhi2015Generalized}. Another network-level neighborhood gap for attributes was proposed as $\sum_i (x_{i,\rm nn}-\langle x\rangle)^2$ in Ref.~\cite{Gehlot2026Quantifying}.

We remark that Evtushenko and Kleinberg~\cite{Evtushenko2023Nodebased} reported some numerical and empirical cases where $\langle\Delta x\rangle_{\rm link}>0$ but $\langle\Delta x\rangle_{\rm ego}<0$, equivalently, where $\rho_{kx}>0$ but $\rho_{\kappa x}<0$. These findings can be understood by the fact that degrees $\{k_i\}$ and weighted degrees $\{\kappa_i\}$ are not necessarily perfectly correlated with each other, when $\rho_{kx}$ and $\rho_{\kappa x}$ may have different signs for some given configurations of attributes as demonstrated in Ref.~\cite{Evtushenko2023Nodebased}, despite the fact that $\rho_{k\kappa}\geq 0$ in Eq.~\eqref{eq:rho_kkappa_result}.

\subsection{Friendship paradoxes in directed, weighted, and signed networks}\label{subsec:direct_weight}

So far we have considered only undirected, binary, and unsigned networks. However, there are a number of real-world systems that can be characterized by networks that are directed, weighted, and/or signed~\cite{Newman2018Networks, Menczer2020First, Diaz-Diaz2026Signed}, enabling us to study FPs and GFPs on those kinds of networks too.

One of the real-world examples for directed social networks can be derived from the online social network (OSN) data~\cite{Kumar2006Structure, Mislove2007Measurement, Penni2017Future}. For example, users on Twitter (currently X) can follow other users who do not necessarily follow back, implying that links between users are directed~\cite{Kwak2010What, Hodas2013Friendship, BenSliman2018Asymmetric}; directed links are also called arcs in graph theory~\cite{Bang-Jensen2009Digraphs}. If a user $i$ follows another user $j$ so that $j$'s posts appear in $i$'s timeline, then $i$ is called $j$'s follower and $j$ is called $i$'s followee or leader. The direction of the link connecting users $i$ and $j$ can be defined as being either $i\to j$ (i.e., $i$ created a link to $j$) or $j\to i$ (i.e., information flows from $j$ to $i$). Which definition is better depends on the context or research question at hand, though we take the latter, following Hodas et al.~\cite{Hodas2013Friendship}. For a directed network, each user or node $i$ is associated with an in-degree $k_i^{\rm in}$, i.e., the number of links pointing to $i$, and an out-degree $k_i^{\rm out}$, i.e., the number of links emanating from $i$. Note that the average in-degree equals to the average out-degree in any network, namely, $\langle k^{\rm in}\rangle=\langle k^{\rm out}\rangle=L/N$, with $L$ denoting the number of directed links. 

The FP-ind in the directed network can be stated in four distinct forms such that your followers/followees have on average more followers/followees than you do~\cite{Hodas2013Friendship}, which can be formally written as
\begin{align}
    k^{\alpha|\beta}_{i,\rm nn}\equiv \frac{1}{k^{\beta}_i}\sum_{j\in \Lambda^{\beta}_i} k^{\alpha}_j>k^{\alpha}_i \quad \textrm{for}\ \alpha,\beta\in\{\rm in, out\},
    \label{eq:kinn_directed_define}
\end{align}
where $\Lambda^{\rm in}_i$ ($\Lambda^{\rm out}_i$) denotes the set of $i$'s in-neighbors ($i$'s out-neighbors), $|\Lambda_i^{\rm in}|=k_i^{\rm in}$, and $|\Lambda_i^{\rm out}|=k_i^{\rm out}$ [see Fig.~\ref{fig:symbols}(b)]. Here, $\beta$ indicates which neighbors to take into account and $\alpha$ does which degrees of those neighbors to average over, so that the comparison is always made against the node's own $k^{\alpha}_i$. Accordingly, four network-level paradox holding probabilities can be computed by
\begin{align}
    H^{\alpha|\beta}\equiv \frac{1}{N}\sum_i \theta(k^{\alpha|\beta}_{i,\rm nn}-k^{\alpha}_i).
    \label{eq:FP_direct_H_define}
\end{align}
For the directed version of the FP-net-link, the average in-degree/out-degree of in-neighbors/out-neighbors of nodes are defined as~\cite{Alipourfard2020Friendship}
\begin{align}
    \langle k^{\alpha|\beta}\rangle_{\rm link} \equiv  \frac{\sum_i k_i^\beta k_i^\alpha}{\sum_i k_i^\beta} = \langle k^\alpha\rangle + \frac{\rho_{k^{\alpha}k^{\beta}}\sigma_{k^{\alpha}}\sigma_{k^{\beta}}}{\langle k^\beta\rangle},
\end{align}
where $\rho_{k^{\alpha}k^{\beta}}$ is the PCC between $\{k_i^{\alpha}\}$ and $\{k_i^{\beta}\}$, and $\sigma_{k^{\alpha}}$ ($\sigma_{k^{\beta}}$) is the standard deviation of $\{k_i^{\alpha}\}$ ($\{k_i^{\beta}\}$). Note that $\rho_{k^{\alpha}k^{\alpha}}=1$ for $\alpha\in\{\rm in,out\}$ and $\langle k^{\rm in|out}\rangle_{\rm link}=\langle k^{\rm out|in}\rangle_{\rm link}$ because $\langle k^{\rm in}\rangle=\langle k^{\rm out}\rangle$ . Thus, whether the directed version of the FP-net-link holds for a nontrivial network with $\sigma_{k^{\alpha}}, \sigma_{k^{\beta}}>0$ is determined by the sign of $\rho_{k^{\alpha}k^{\beta}}$. Next, the degree-dependent paradox holding probabilities are defined as follows:
\begin{align}
    h_{\beta}(k^{\alpha}) \equiv \frac{1}{|V_{k ^{\alpha}}|}\sum_{i\in V_{k^{\alpha}}} \theta(k^{\alpha|\beta}_{i,\rm nn}-k^{\alpha}),
    \label{eq:FP_hk_direct_define}
\end{align}
where $V_{k^{\alpha}}$ is the set of nodes having the degree $k^{\alpha}$ for $\alpha\in\{\rm in, out\}$. Also, the degree-dependent FPs for the set of nodes with degree $k^\alpha$ are written as
\begin{align}
    \langle k^{\alpha|\beta}\rangle_{\rm nn}(k^{\alpha})\equiv \frac{1}{|V_{k^{\alpha}}|}\sum_{i\in V_{k^{\alpha}}}k^{\alpha|\beta}_{i,\rm nn} > k^{\alpha}.
    \label{eq:k_nn_k_direct}
\end{align}
The FP-net-ego for directed networks can be similarly defined and studied, so can directed versions of GFP-ind, GFP-net-link, GFP-net-ego, and so on. 

Here we introduce directed versions of network-level GFPs studied in Refs.~\cite{BenSliman2018Asymmetric, Alipourfard2020Friendship}, for which one can first define for $\alpha\in\{\rm in,out\}$
\begin{align}
    \label{eq:x_link_directed}
   \langle x\rangle^{\alpha}_{\rm link} \equiv \frac{\sum_i k_i^{\alpha} x_i}{\sum_i k_i^{\alpha}} = \langle x\rangle + \frac{\rho_{k^{\alpha}x}\sigma_{k^{\alpha}}\sigma_x}{\langle k^\alpha\rangle}
\end{align}
with $\rho_{k^{\alpha}x}$ denoting the PCC between $\{k^{\alpha}_i\}$ and $\{x_i\}$ of nodes, and
\begin{align}
   \label{eq:x_ego_directed}
   &\langle x\rangle^{\alpha}_{\rm ego} \equiv \frac{1}{N}\sum_i \frac{1}{k_i^{\alpha}}\sum_{j\in \Lambda_i^{\alpha}}x_j = \langle x\rangle + \rho_{\kappa^{\alpha}x}\sigma_{\kappa^{\alpha}}\sigma_x,
\end{align}
where we respectively define the weighted in-degrees and out-degrees\footnote{Similarly to the weighted degree defined by Cantwell et al.~\cite{Cantwell2021Friendship}, one can define the weighted in-degrees and out-degrees in directed networks; for example, $\kappa^{\rm in}_i$ is defined as the sum of inverse in-degrees of $i$'s out-neighbors by considering that $\langle x\rangle^{\rm in}_{\rm ego} =\frac{1}{N}\sum_{ij} \frac{1}{k_i^{\rm in}} a_{ji}x_j =\frac{1}{N}\sum_{j} (\sum_i \frac{1}{k_i^{\rm in}} a_{ji})x_j =\frac{1}{N}\sum_{j}\kappa^{\rm in}_j x_j$, where $\kappa^{\rm in}_j\equiv \sum_{i\in \Lambda_j^{\rm out}} \frac{1}{k^{\rm in}_i}$.} as
\begin{align}
    &\kappa^{\rm in}_i \equiv \sum_{j\in\Lambda^{\rm out}_i} \frac{1}{k^{\rm in}_j},\\
    &\kappa^{\rm out}_i \equiv \sum_{j\in\Lambda^{\rm in}_i} \frac{1}{k^{\rm out}_j}.
\end{align}
In addition, $\rho_{\kappa^{\alpha}x}$ is the PCC between $\{\kappa^{\alpha}_i\}$ and $\{x_i\}$, and $\sigma_{\kappa^{\alpha}}$ is the standard deviation of $\{\kappa_i^{\alpha}\}$. In particular, Alipourfard et al.~\cite{Alipourfard2020Friendship} considered binary attributes with $x_i\in\{0,1\}$, and focused on the neighborhood gaps for the attribute:
\begin{align}
    &\langle \Delta x\rangle^{\rm out}_{\rm link}\equiv \langle x\rangle^{\rm out}_{\rm link}-\langle x\rangle,\\ 
    &\langle \Delta x\rangle^{\rm in}_{\rm ego}\equiv \langle x\rangle^{\rm in}_{\rm ego}-\langle x\rangle,
\end{align}
which they called global and local perception biases, respectively. They identified conditions for $\langle \Delta x\rangle^{\rm in}_{\rm ego}\geq \langle \Delta x\rangle^{\rm out}_{\rm link}\geq 0$ as well as for the cases when these neighborhood gaps have different signs.

As for the undirected, weighted networks~\cite{Newman2003Structure, Barrat2004Architecture}, the link connecting nodes $i$ and $j$ is assigned a link weight $w_{ij}$. For example, $w_{ij}$ can be the total number or duration of phone calls between two phone users $i$ and $j$~\cite{Onnela2007Structure} or the weighted number of papers coauthored by two researchers $i$ and $j$~\cite{Barrat2004Architecture}. For the undirected network, $w_{ij}=w_{ji}$. The strength of a node $i$ is defined as the sum of weights of links adjacent to $i$~\cite{Menczer2020First}:
\begin{align}
    s_i\equiv \sum_{j\in \Lambda_i} w_{ij}.
    \label{eq:s_i_define}
\end{align}
The individual-level FP can be stated such that on average your neighbors have higher strengths than you do:
\begin{align}
    \frac{1}{k_i}\sum_{j\in \Lambda_i} s_j>s_i.
    \label{eq:FP_weighted_define}
\end{align}
We observe that the above definition of the FP on the weighted network looks similar to the GFP on the binary network in Eq.~\eqref{eq:GFP_ind_define}, except for node strengths replacing nodal attributes in Eq.~\eqref{eq:GFP_ind_define}. Therefore, it is expected that whether the FP-net-link in the weighted network holds or not depends on the sign of the PCC between degrees and strengths of nodes, denoted by $\rho_{ks}$. A slightly different definition of the FP on the weighted network was proposed in Ref.~\cite{Evtushenko2024Comprehensive}, where $k_i$ on the left hand side of Eq.~\eqref{eq:FP_weighted_define} is replaced by $s_i$.

For the directed, weighted network, the link weight is not necessarily symmetric, that is, $w_{ij}\neq w_{ji}$ in general. Then the in-strength and out-strength of the node $i$ are respectively defined as~\cite{Menczer2020First, Berenhaut2019Friendship}
\begin{align}
    &s^{\rm in}_i\equiv \sum_{j\in \Lambda^{\rm in}_i} w_{ji},\ \\ 
    &s^{\rm out}_i\equiv \sum_{j\in \Lambda^{\rm out}_i} w_{ij}.
    \label{eq:s_i_in_out_define}
\end{align}
The individual-level FP on the directed, weighted network can be formulated as
\begin{align}
    \frac{1}{k^{\beta}_i}\sum_{j\in \Lambda^{\beta}_i} s^{\alpha}_j>s^{\alpha}_i
\end{align}
for $\alpha,\beta\in\{\rm in, out\}$. The network-level FPs and their generalizations on the directed, weighted network can be similarly defined and studied. 

Finally, links in a signed network have signs, i.e., positive or negative~\cite{Diaz-Diaz2026Signed}, which can be seen as a two-layer network~\cite{Jo2006Immunization} but having one positive and one negative layers. Ghasemian and Christakis~\cite{Ghasemian2023Enmity} defined the enmity paradox on the negative layer, stating that your enemies have on average more enemies than you do. To be precise, let $k^{+}_i$ ($k^{-}_i$) denote the number of positive (negative) links adjacent to the node $i$. For $\alpha,\beta\in\{+,-\}$, one can formulate the individual-level FP in the signed network as
\begin{align}
    \frac{1}{k^{\beta}_i}\sum_{j\in \Lambda^{\beta}_i} k^{\alpha}_j>k^{\alpha}_i,
    \label{eq:kinn_signed_define}
\end{align}
where $\Lambda^{+}_i$ ($\Lambda^{-}_i$) denotes the set of $i$'s friends (enemies). The case with $\alpha=\beta=-$ indicates the enmity paradox at the individual level, while the case with $\alpha=\beta=+$ corresponds to the original, individual-level FP. The other cases with $\alpha\neq\beta$ imply that your enemies (friends) have on average more friends (enemies) than you do. Network-level FPs and GFPs in signed networks were also studied in Ref.~\cite{Ghasemian2023Enmity}.

\subsection{Perception methods}\label{subsec:percept}

Taking an average of degrees of a node $i$'s neighbors [$k_{i,\rm nn}$ in Eq.~\eqref{eq:k_inn_define}] is not the only way of perceiving or summarizing the $i$'s neighborhood, where the $i$'s neighborhood indicates the set of degrees of $i$'s neighbors, i.e., $K_i\equiv \{k_j|j\in \Lambda_i\}$. Alternatively, one can take the median of degrees for the comparison to the focal node's degree~\cite{Feld1991Why, Ugander2011Anatomy, Kooti2014Network, Momeni2016Qualities, Lee2019Impact}, enabling us to write the median-based FP-ind as
\begin{align}
    k_{i, \rm md}>k_i,
\end{align}
where $k_{i,\rm md}$ denotes the median degree of $K_i$. The median-based FP has been called a strong FP~\cite{Kooti2014Network, Lerman2025Strong} as the median is more robust than the average to very high degrees in $K_i$, particularly in the network showing a heavy-tailed degree distribution. Lee et al.~\cite{Lee2019Impact} introduced another perception method based on the fraction of neighbors whose degree is greater than the degree of the focal node. To compare these three perception methods more systematically, three versions of ``peer pressure" for each node $i$, namely, mean-based, median-based, and fraction-based peer pressures are respectively defined as
\begin{align}
    &h_{i,\rm mn}\equiv \theta\left(k_{i,\rm nn} -k_i\right),\\ 
    &h_{i,\rm md}\equiv \theta\left(k_{i,\rm md} -k_i\right),\\ 
    &h_{i,\rm fr}\equiv \frac{1}{k_i}\sum_{j\in \Lambda_i}\theta(k_j -k_i).\label{eq:h_i_fr_define}
\end{align}
Notably, both $h_{i,\rm mn}$ and $h_{i,\rm md}$ can have a value of either $0$ or $1$, whereas $h_{i,\rm fr}$ can have a value in the range of $[0,1]$. Note that Zhou et al.~\cite{Zhou2020Revisiting} focused on the number of the node $i$'s neighbors whose degrees are higher than $k_i$, i.e., $\sum_{j\in \Lambda_i}\theta(k_j -k_i)$. The network-level peer pressures, respectively denoted by $H_{\rm mn}$, $H_{\rm md}$, and $H_{\rm fr}$, are then calculated as the network-level averages of the corresponding individual peer pressures~\cite{Lee2019Impact}. Here $H_{\rm mn}$ is identical to $H$ in Eq.~\eqref{eq:FP_H_define}\footnote{Recently, Lee~\cite{Lee2026Friendshipparadox} independently proposed two network-level quantities $\phi_{\rm global}$ and $\phi_{\rm local}$, which correspond to $H_{\rm mn}$ and $H_{\rm md}$, respectively.}.

These three versions of the neighborhood perception, i.e., mean-based, median-based, and fraction-based peer pressures were generalized to the following formulation~\cite{Lee2019Impact}:
\begin{equation}
    \label{eq:FP_h_general}
    h_i \equiv \psi\left[\frac{1}{k_i}\sum_{j\in \Lambda_i}\phi(k_j -k_i)-\xi \right],
\end{equation}
where $\xi$ is some intrinsic value for nodes. One recovers the mean-based peer pressure if $\psi(x)=\theta(x)$, $\phi(x)=x$, and $\xi=0$. The median-based case is recovered if $\psi(x)=\theta(x)$, $\phi(x)=\theta(x)$, and $\xi=1/2$. Finally, the fraction-based case is recovered when $\psi(x)=x$, $\phi(x)=\theta(x)$, and $\xi=0$. Then, other perception methods can be studied by specifying $\psi(x)$, $\phi(x)$, and $\xi$ in Eq.~\eqref{eq:FP_h_general} as appropriate. Similarly, one can adopt the same formulation for the GFP as follows:
\begin{equation}
    \label{eq:GFP_h_general}
    h_{{\rm at},i} \equiv \psi\left[\frac{1}{k_i}\sum_{j\in \Lambda_i}\phi(x_j -x_i)-\xi \right].
\end{equation}
We remark that individuals may perceive their neighborhoods in different ways, meaning that the functional forms of $\psi(x)$ and $\phi(x)$, and the value of $\xi$ can vary across individuals.

Different perception methods might have different impacts on the FP itself as well as on dynamical processes driven by the FP. For example, Lee et al.~\cite{Lee2019Impact} studied the impact of perception methods on the voter model~\cite{Castellano2009Statistical, Starnini2026Opinion} driven by the FP; in the model, each node $i$ replaces its opinion by that of its randomly chosen neighbor and the replacement probability is given as $\epsilon + (1-\epsilon) h_i$ with a small positive constant $\epsilon$ and $h_i$ denoting the $i$'s peer pressure. The time for the system to reach the consensus turns out to depend on whether the perception method is mean-based, median-based, or fraction-based, in other words, how people perceive their neighborhoods.

\section{Results on model networks}\label{sec:model_net}

In this section, we review the literature on the FP and its generalizations studied on the concrete network structure using a variety of network models. As mentioned, real-world social networks have been characterized by various network measures or quantities, such as the degree distribution, degree assortativity, clustering coefficient, average path length, and community structure among others~\cite{Barabasi1999Emergence, Jeong2003Complex, Stumpf2012Critical, Broido2019Scalefree, Voitalov2019Scalefree, Newman2002Assortative, Newman2003Mixinga, Watts1998Collective, Milgram1967Smallworld, Fortunato2010Community, Schaub2017Many, Javed2018Community, Fortunato202220, Granovetter1973Strength, Onnela2007Structure, Onnela2007Analysis, Newman2003Structure, Jo2018Stylized}. The relevant question is how such network properties affect the FP and its generalizations on all levels from individual nodes to the entire network. Because most empirical networks are consequences of the interplay of multiple factors affecting the different aspects of networks, a systematic approach is required to scrutinize the effects of each factor and their combinations. For this, we need network models by which one can control different factors affecting the network properties. Network scientists have developed and studied various network models including Erd\H{o}s--R\'enyi (ER) random graphs~\cite{Erdos1960Evolution}, Watts--Strogatz (WS) small-world networks~\cite{Watts1998Collective}, Barab\'asi--Albert (BA) scale-free networks~\cite{Barabasi1999Emergence}, and configuration models~\cite{Molloy1995Critical, Newman2001Random} to name a few. Below we give an overview of the analytical and numerical results on the model networks by focusing on (i) paradox holding probabilities and (ii) friendship indexes and neighborhood gaps.

Before going further, we introduce the degree assortativity considering its important role in understanding the FPs and GFPs on the model and real networks. The degree assortativity quantifies the similarity between degrees of neighboring nodes, and it is defined as~\cite{Newman2002Assortative, Newman2003Mixinga}
\begin{align}
    \rho_k\equiv \frac{L\sum_l k_lk'_l-\left[\sum_l\frac{1}{2}(k_l+k'_l)\right]^2} {L\sum_l\frac{1}{2}({k_l}^2+{k'_l}^2) -\left[\sum_l\frac{1}{2}(k_l+k'_l)\right]^2},
    \label{eq:assort_define_data}
\end{align}
where $k_l$ and $k'_l$ are degrees of two nodes connected by the $l$th link ($l=1,\ldots,L$). Equivalently, it is written as
\begin{align}
    \rho_k \equiv \frac{1}{\sigma^2_q}\sum_{kk'} kk'(e_{kk'}-q_kq_{k'}),
    \label{eq:assort_define}
\end{align}
where $e_{kk'}$ is the joint probability distribution of excessive degrees $k$ and $k'$ of two nodes connected by a randomly chosen link, $q_k=\sum_{k'} e_{kk'}$ is the excessive degree distribution, and $\sigma^2_q$ is the variance of excessive degrees. The value of $\rho_k$ ranges from $-1$ to $1$. The positive $\rho_k$ implies the tendency of connecting nodes with similar degrees, implying that the network is assortative or homophilic~\cite{McPherson2001Birds}. In contrast, the negative $\rho_k$ indicates the opposite tendency, implying the disassortative or heterophilic network, while the case of $\rho_k=0$ indicates no correlations between degrees of neighboring nodes.

\subsection{Paradox holding probabilities}\label{subsec:net_paradox_prob}

Hazra et al.~\cite{Hazra2026Friendship} computed the network-level paradox holding probability $H$ in Eq.~\eqref{eq:FP_H_define} on the ER random graph, the configuration model with power-law degree distribution, and a variant of the BA scale-free network, all in the sparse regime. For the case with the ER graph with the connection probability $p$, the value of $H$ decreases from $1$ to $1/2$ as the average degree $\langle k\rangle =(N-1)p$ increases from $0$ to $\infty$. They also considered the configuration model with the power-law degree distribution $P(k)=k^{-\alpha}/\zeta(\alpha)$, where $\alpha>2$ and $\zeta(\cdot)$ is the Riemann zeta function, and found that $H$ approaches $1$ both for $\alpha\to 2$ and for $\alpha\to\infty$, while it is less than $1$ for the intermediate range of $\alpha$. Finally, they considered a growing network with a preferential attachment rule; at each timestep a new node is added with one link, and it chooses one of the existing nodes, say $i$, with a probability proportional to $k_i+\delta$ ($\delta\geq -1$) to make a connection to $i$. This network shows a power-law degree distribution with an exponent $3+\delta$. They found that as $\delta$ increases from $-1$, $H$ first decreases from $1$ and then increases to approach $1$.

Lee et al.~\cite{Lee2019Impact} numerically studied the effects of the degree distribution $P(k)$ and the degree assortativity $\rho_k$ in Eq.~\eqref{eq:assort_define} on the mean-based, median-based, and fraction-based paradox holding probabilities, namely, $H_{\rm mn}$, $H_{\rm md}$, and $H_{\rm fr}$ (Sec.~\ref{subsec:percept}). When $P(k)$ is modeled by an exponential function on the discrete domain, as $\rho_k$ increases, $H_{\rm mn}$ increases and then decreases, whereas $H_{\rm fr}$ monotonically decreases and $H_{\rm md}$ stays almost flat before dropping sharply. This result implies that how individuals perceive their neighborhoods affects the FP behaviors. They also reported that all of mean-based, median-based, and fraction-based $h(k)$s are increasing and then decreasing as $k$ increases for assortative networks ($\rho_k>0$), where nodes tend to be connected to others with similar degrees\footnote{The similar result for the mean-based FP was reported in Ref.~\cite{Jo2014Generalized}.}. In this case, intermediate-degree nodes experience the FP most severely, compared to both low-degree and high-degree nodes. It is because intermediate-degree nodes typically have both low-degree and high-degree neighbors, while the latter tends to dominate the average degree of neighbors. For disassortative networks ($\rho_k<0$), where high-degree nodes tend to be connected to low-degree nodes, $h(k)$ monotonically decreases with $k$ for every perception method.

Wu et al.~\cite{Wu2017NeighborNeighbor} studied the impact of the degree assortativity $\rho_k$ on the median-based FP in terms of the degree-dependent paradox holding probability $h_{\rm md}(k)$ and the network-level paradox holding probability $H_{\rm md}$. For large $k$, they derived
\begin{align}
    h_{\rm md}(k)\approx 1-\Phi\left[\frac{(1-2\mu_k)\sqrt{k}}{2\sqrt{\mu_k(1-\mu_k)}}\right],
\end{align}
where $\Phi$ is the cumulative distribution of the normal distribution, and $\mu_k$ is the probability that a random neighbor of the node with degree $k$ has a higher degree than $k$, precisely,
\begin{align}
    \mu_k=\sum_{k'>k}\frac{e_{kk'}}{q_k}.
\end{align}
They adopted the bivariate log-normal distribution for $e_{kk'}$, which has three parameters, i.e., the mean $m$ and variance $s^2$ of $\log k$, and the correlation parameter $r$ between $k$ and $k'$, to get
\begin{align}
    \mu_k=1-\Phi\left(\frac{\log k-m}{s}\sqrt{\frac{1-r}{1+r}}\right).
\end{align}
The degree assortativity $\rho_k$ is related to $r$ such that $\rho_k=(e^{rs^2}-1)/(e^{s^2}-1)$, where $\rho_k$ and $r$ have the same sign. They observed that $h_{\rm md}(k)$ is a monotonically decreasing function of $k$ for both positive and negative values of $\rho_k$, and that $H_{\rm md}$ is monotonically decreasing with $\rho_k$. In a similar setup but with different functions of $P(k)$, Lee et al.~\cite{Lee2019Impact} found the same monotonic behavior of $H_{\rm md}$, but showed that $h_{\rm md}(k)$ is an increasing and then decreasing function of $k$ when $\rho_k>0$.

The paradox holding probability for the GFP was analyzed for the network in which nodes' attributes are given from some distribution $P(x)$ for continuous $x$. Since the transformation of $x_i$ to $x_i+c$ with some constant $c$ does not affect the GFP behavior in Eq.~\eqref{eq:GFP_ind_define}, the range of $0\leq x<\infty$ is assumed without loss of generality. Jo and Eom~\cite{Jo2014Generalized} defined the paradox holding probability for a node with degree $k$ and attribute $x$ as
\begin{align}
    &h(k,x)\equiv \Pr\left(\frac{1}{k}\sum_{j=1}^k x_j >x\right)= \left(\prod_{j=1}^k\int_0^\infty {\rm d}x_j\right) \notag\\
    &\times P(x_1,\ldots,x_k|x)\theta\left(\frac{1}{k}\sum_{j=1}^k x_j-x\right),
    \label{eq:hkx_define}
\end{align}
where $x_j$s are attributes of $k$ neighbors of the focal node, and $P(x_1,\ldots,x_k|x)$ is the joint probability distribution of $x_j$s conditioned on the focal node's attribute $x$. The simplest case is such that all $x$ values are fully uncorrelated with each other, enabling to factorize the joint probability distribution as $P(x_1,\ldots,x_k|x)=\prod_{j=1}^k P(x_j)$. Then one gets the Laplace transform of $h(k,x)$ with respect to $x$ as
\begin{align}
    \tilde h(k,s)=\frac{1}{s}\left[1-\tilde P\left(\frac{s}{k}\right)^k\right],
\end{align}
where $\tilde P(s)$ is the Laplace transform of $P(x)$. For the demonstration of the analytical result, they used the Gamma distribution $P(x)=x^{\alpha-1}e^{-x/\beta}/[\beta^\alpha\Gamma(\alpha)]$, where $\Gamma(z)\equiv \int_0^\infty {\rm d}t t^{z-1}e^{-t}$ is a Gamma function, leading to the analytical result of $h(k,x)$ as
\begin{align}
    h(k,x)=\frac{\Gamma(\alpha k,\alpha k\frac{x}{\langle x\rangle})}{\Gamma(\alpha k)},
\end{align}
where $\Gamma(s,z)\equiv \int_z^\infty {\rm d}t t^{s-1}e^{-t}$ is the upper incomplete Gamma function and $\langle x\rangle=\alpha\beta$~\cite{Jo2014Generalized}. Even in such a simplest case, the behavior of $h(k,x)$ is nontrivial; e.g., when $x>\langle x\rangle$, $h(k,x)$ is an increasing and then decreasing function of $k$, which is in contrast to the naive expectation that nodes with higher $k$ and $x$ will be less likely to experience the GFP. Then, the cases with correlated attributes of neighboring nodes were studied by numerical simulations. For this, the attribute assortativity $\rho_x$ is defined as the PCC between attributes of neighboring nodes. They found that when $\rho_x>0$, nodes with relatively high degrees and attributes have larger values of $h(k,x)$ than in the uncorrelated case ($\rho_x=0$). It is because those nodes have neighbors who have even higher degrees and attributes than their own degrees and attributes.

Later Jo et al.~\cite{Jo2021Analytical, Jo2022Copulabased} adopted the copula methods to derive analytical results of $h(k,x)$ for networks with the correlation between attributes of neighboring nodes. In essence, the copula method enables us to write down the explicit form of the joint probability distribution of two or more correlated random variables~\cite{Nelsen2006Introduction}. The copula method has been used in various disciplines such as finance~\cite{Embrechts2009Copulas, Cossette2013Multivariate}, time series analysis~\cite{Jo2019Analytically, Jo2019Copulabased, Yu2025Analysis, Lee2026Copulabased}, astronomy~\cite{Takeuchi2010Constructing, Takeuchi2020Constructing}, meteorology~\cite{Yue2002Bivariate}, biology~\cite{Ray2020CODC}, and intelligent systems~\cite{Horvath2020CopulaBased}. Among many other copulas, the Farlie--Gumbel--Morgenstern (FGM) copula~\cite{Schucany1978Correlation, Huang1984Correlation, Nelsen2006Introduction} turns out to be relatively simple but general enough to study the GFP. Using the FGM copula, the joint probability distribution of attributes of two neighboring nodes reads
\begin{align}
    P(x,x')=P(x)P(x')[1+rf(x)f(x')],
\end{align}
where a correlation parameter $r\in[-1,1]$ controls the degree of correlation between $x$ and $x'$, and $f(x)\equiv 2F(x)-1$, where $F(x)\equiv \int_0^x{\rm d}xP(x)$ is the cumulative probability distribution of $P(x)$. The parameter $r$ and the attribute assortativity $\rho_x$ are linear with each other:
\begin{align}
    \rho_{x}=\frac{r}{\sigma_x^2}\left[\int_0^\infty {\rm d}x x P(x)f(x)\right]^2\equiv cr.
    \label{eq:rhoxx_r}
\end{align}
The upper bound of the value of $c$ for any functional form of $P(x)$ was proven to be $1/3$, indicating that $|\rho_{x}|\le 1/3$~\cite{Schucany1978Correlation}.

The first application of the FGM copula to the GFP is based on the assumption that the network is tree-like~\cite{Jo2021Analytical}, implying that the possible correlation between attributes of the focal node's neighbors can be ignored. Then $P(x_1,\ldots,x_k|x)$ in Eq.~\eqref{eq:hkx_define} can be written as
\begin{align}
    P(x_1,\ldots,x_k|x)=\prod_{j=1}^k P(x_j|x)=\prod_{j=1}^k P(x_j)[1+rf(x_j)f(x)].
\end{align}
Assuming that $P(x)=\lambda e^{-\lambda x}$ with $\lambda>0$ and that $|r|\ll 1$, one obtains up to the first order of $r$
\begin{align}
  h(k,x) = h_0(k,x) + r h_1(k,x) +\mathcal{O}(r^2),
  \label{eq:hkx_mean_expo}
\end{align}
where
\begin{align}
    &h_0(k,x)\equiv g(k,\lambda kx), \\  
    &h_1(k,x)\equiv k(-1)^{k+1} e^{-2\lambda kx} (1-2e^{-\lambda x})\left[ g(k,-\lambda kx)-1 \right].
\end{align}
Here $g(s,z)\equiv \Gamma(s,z)/\Gamma(s)$ is the normalized upper incomplete Gamma function. This analytical solution is in good agreement with numerical simulations for sufficiently small values of $|r|$. Since $h_1(k,x)>0$ for all $k$ and $x$, the positive correlation between attributes of neighboring nodes ($r>0$) makes the region of large $h(k,x)$ expand into the area of high $k$ and $x$. It implies that nodes with high $k$ and $x$ are more likely to experience the GFP than they would otherwise, because those nodes have neighbors with even higher $k$ and $x$. Using the same framework, the median-based and fraction-based cases were also analytically studied in Ref.~\cite{Jo2021Analytical}.

The second application of the FGM copula considers the GFP on clustered networks, in which the triangles are abundant~\cite{Jo2022Copulabased}. The vine copula formulation~\cite{Takeuchi2020Constructing} enables to write $P(x_1,\ldots,x_k|x)$ in Eq.~\eqref{eq:hkx_define} as
\begin{align}
    &P(x_1,\ldots,x_k|x)= \prod_{j=1}^k P(x_j)[1+rf(x_j)f(x)]\notag\\
    &\times \prod_{\{jj'|a_{jj'}=1\}}[1+rf(x_j)f(x_{j'})].
\end{align}
Here $a_{jj'}=1$ means that two neighbors $j$ and $j'$ of the focal node are connected to each other, implying triangles adjacent to the focal node. Assuming that $P(x)=\lambda e^{-\lambda x}$ with $\lambda>0$ and that $|r|\ll 1$, one obtains up to the first order of $r$
\begin{align}
  h(k,x) = h_0(k,x) + r h_1(k,x)+ rn h_2(k,x) +\mathcal{O}(r^2),
  \label{eq:hkx_mean_expo_vine}
\end{align}
where $n$ is the number of links between neighbors of the focal node, and
\begin{align}
    &h_2(k,x)\equiv (-1)^k e^{-2\lambda kx} \left\{(1-2\lambda kx)\left[g(k,-\lambda kx)-1\right]\right.\notag\\
    &\left. -2k \left[g(k+1,-\lambda kx)-1\right]\right\}.
\end{align}
Here $h_2(k,x)$ is found to be positive (negative) when $x>\langle x\rangle$ ($x<\langle x\rangle$). It implies that when your attribute is higher (lower) than the average, the more connected your neighbors are to each other, the more (less) likely you are to experience the GFP. The above analytical solution is in good agreement with numerical simulations for sufficiently small values of $|r|$. For the numerical simulations, the substrate networks were generated using the configuration model with the given sequence of the number of single links and the number of triangles adjacent to each node, which was independently developed by Newman~\cite{Newman2009Random} and Miller~\cite{Miller2009Percolation}. On the top of the substrate network, the randomly drawn attributes from $P(x)$ are assigned to nodes and then shuffled until the desired attribute assortativity $\rho_x$ is reached.

\subsection{Friendship indexes and neighborhood gaps}\label{subsec:net_FI_gap}

We now discuss friendship indexes and neighborhood gaps studied in several model networks. Pal et al.~\cite{Pal2019Study} introduced the friendship index for each node in Eq.~\eqref{eq:FI_define} to propose the network-level measures by taking an arithmetic mean, geometric mean, and harmonic mean of individual friendship indexes over all nodes in the network, respectively defined as
\begin{align}
    &\langle I\rangle \equiv \frac{1}{N}\sum_i I_i, \label{eq:arith_mean_FI_define} \\
    &\langle \log I\rangle \equiv \frac{1}{N}\sum_i \log I_i,\\
    &\langle I^{-1}\rangle^{-1} \equiv \left(\frac{1}{N}\sum_i I^{-1}_i\right)^{-1}.
    \label{eq:harmonic_mean_FI_define}
\end{align}
These means were calculated exactly for regular random graphs, star graphs, and complete bipartite graphs. In addition, the following has been proven: (i) $\langle I\rangle$ has a lower bound of $1$ for all networks, (ii) $\langle I\rangle$ has a maximum value for the intermediate range of $c\equiv Np$ for the ER graphs when $N\to\infty$, where $p$ is the probability that two randomly chosen nodes are connected with each other, and (iii) $\langle I\rangle$ for the BA scale-free network diverges as the network size increases with the time step. Pal et al.~numerically studied the behaviors of $\langle I\rangle$ and other means for the ER, WS, and BA networks to find consistent behaviors with the analytical results.

Sidorov et al.~\cite{Sidorov2021Friendship} and Grigoriev et al.~\cite{Grigoriev2024Dynamics} studied the dynamics of the friendship index of nodes in the growing process of the BA networks with linear and nonlinear preferential attachment rules~\cite{Krapivsky2000Connectivity} as well as in the growing process of the Holme--Kim model with triadic closure~\cite{Holme2002Growing}.

Fotouhi et al.~\cite{Fotouhi2015Generalized} and Momeni and Rabbat~\cite{Momeni2015Measuring} studied a growing network model with preferential attachment. In the model, each node is assigned an attribute $x$ drawn from $P(x)$ when the node is added to the network, and this attribute remains constant. The newly added node at each time step $t$ randomly chooses an existing node, say $i$, with probability proportional to $k_i+x_i$, to make a connection to $i$. They derived the joint probability distribution $P(k,x)$ in the steady state for $t\to\infty$. Using such $P(k,x)$, they calculated $\langle \Delta k\rangle_{\rm link}$ in Eq.~\eqref{eq:FP_gap_link_define} for the FP as well as $\langle\Delta x\rangle_{\rm link}$ in Eq.~\eqref{eq:GFP_gap_link_define} for the GFP.

Cantwell et al.~\cite{Cantwell2021Friendship} derived the distributions of the neighborhood gap in Eq.~\eqref{eq:deltak_i} for the ER graph, the configuration model with the truncated power-law degree distribution, and the random network with the truncated power-law degree distribution as well as with the controllable degree assortativity. For the ER graph, they obtained the neighborhood gap distribution as
\begin{align}
    P(\Delta k)=\sum_k P(k)P(\Delta k|k),
\end{align}
where
\begin{align}
    &P(k)=\frac{\langle k\rangle^k}{k!}e^{-\langle k\rangle},\\
    &P(\Delta k|k)=\frac{(\langle k\rangle k)^{k\Delta k+k^2-k}}{(k\Delta k+k^2-k)!}e^{-\langle k\rangle k}.
\end{align}
This gives the average and variance of $\Delta k$ as $1$ and $\langle k\rangle (1+\langle k^{-1}\rangle)$, respectively. As $\langle k\rangle$ increases, $P(\Delta k>0)$, or equivalently the network-level paradox holding probability $H$ in Eq.~\eqref{eq:FP_H_define}, turns out to decrease to $1/2$. We remark that $P(\Delta k >0|k)$ can be related to the degree-dependent paradox holding probability $h(k)$ in Eq.~\eqref{eq:FP_hk_define}. In addition, $P(\Delta k)$ was calculated for the random network with the truncated power-law degree distribution $P(k)\propto k^{-\alpha}e^{-\beta k}$ and the degree assortativity $\rho_k$. It is found that the average of $\Delta k$ monotonically decreases with $\rho_k$, whereas $H$ is an increasing and then decreasing function of $\rho_k$ for the range of $[-0.7,0.7]$. A similar non-monotonic behavior of $H$ was reported in Ref.~\cite{Lee2019Impact}.

Hazra et al.~\cite{Hazra2026Friendshipa} studied the FP in finite and infinite trees. When the numbers of nodes with positive, negative, and zero $\Delta k_i$ in Eq.~\eqref{eq:deltak_i} are respectively denoted by $n_+$, $n_-$, and $n_0$, they proved that $\Delta n\equiv n_+-n_-\geq 0$ for the arbitrary finite tree, being translated to $H\gtrsim 1/2$ when $n_0$ is negligible, and derived the lower bound of $\Delta n$. For the infinite tree, they identified the condition for $\Delta n\geq 0$.

Bhattacharya et al.~\cite{Bhattacharya2026Triangle} generalized the FP by considering the number of triangles attached to the node as an attribute $x_i$. The triangle FP can be stated that your friends are on average more involved in triangles than you are. They derived the analytical expressions of $\langle \Delta x\rangle_{\rm ego}$ in Eq.~\eqref{eq:GFP_gap_ego_define} on finite deterministic graphs and ER graphs for both sparse and dense regimes.

\section{Empirical results}\label{sec:empiric} 

Since Scott L. Feld's first demonstration of the FP using the high school students' network~\cite{Feld1991Why}, various social network data, whether small or big, online or offline, have been collected and analyzed to reveal a much richer picture of the FP and its generalizations. These empirical findings are often beyond what model networks can describe or predict, requiring us to devise more realistic models and refined theory too.

\subsection{Friendship paradoxes}\label{subsec:empirical_FP}

The first empirical demonstration of the FP is found in the very paper that introduced the FP. Feld~\cite{Feld1991Why} used data on reciprocated friendships among adolescent girls in Marketville, which is one of the twelve high schools surveyed in James Coleman's \textit{The Adolescent Society} (1961). He exemplified the individual-level FP in Eq.~\eqref{eq:FP_ind_define} by computing the values of $k_i$ and $k_{i,\rm nn}$ of eight girls. Analysis of the full network of girls in Marketville showed that 80 among 146 girls with at least one mutual friend have fewer friends than the average degrees of their friends, while 41 girls have more friends than their friends. That is, the paradox holding probability in Eq.~\eqref{eq:FP_H_define} is $H=80/146 \approx 0.55$. He also discussed that the paradox is affected by the degree assortativity, that the same eight girls experienced the FP when using the median of their friends' degrees, and that the same phenomenon was observed for Marketville boys as well as girls and boys in the other high schools in Coleman's mentioned book.

Two decades later, Ugander et al.~\cite{Ugander2011Anatomy} confirmed the FP by analyzing a large-scale online social network data, that is, the entire graph of active Facebook users as of May~2011 with $N \approx 721$M users and $L \approx 68.7$B reciprocal friendships between them. They obtained that $\langle k\rangle_{\rm link}\approx 635$ and $\langle k\rangle\approx 190$, implying the FP-net-link in Eq.~\eqref{eq:FP_net_link_define}. They found that $H=92.7\%$ of users have fewer neighbors than the average degree of their neighbors, while $H_{\rm md}=83.6\%$ of users experience the median-based FP. The average degree of neighbors of nodes with degree $k$, i.e., $\langle k\rangle_{\rm nn}(k)$ in Eq.~\eqref{eq:k_nn_k}, turned out to be larger than $k$ up to $k\approx 700$, covering most users in the data. It is consistent with the expectation that small-degree nodes are more likely to experience the FP than high-degree ones. However, the most common degree among one's neighbors equaled one's own degree for $k\leq 120$, requiring us to be more careful when interpreting the FP-ind for low- to intermediate-degree nodes.

Jo and Eom~\cite{Jo2014Generalized} explained the degree-dependent paradox holding probability $h(k)$ in Eq.~\eqref{eq:FP_hk_define} measured for the Physical Review (PR) and Google Scholar (GS) coauthorship networks in Ref.~\cite{Eom2014Generalized}. Here $h(k)$ increases and then decreases with $k$ for the PR network, while it monotonically decreases with $k$ for the GS network. Computing $h(k)$ for model networks with tunable degree assortativity, they showed that the increasing behavior of $h(k)$ for small $k$ appears only when the network is assortative. It accounts for the contrasting behaviors between the PR network ($\rho_k\approx 0.47$) and the GS network ($\rho_k\approx -0.02$); degrees of neighboring nodes in the GS network are essentially uncorrelated.

Wu et al.~\cite{Wu2017NeighborNeighbor} measured the median-based paradox holding probability $H_{\rm md}$ for six real-world networks, i.e., LiveJournal user network, YouTube user network, Skitter internet graph, Google hyperlink graph, arXiv HEP citation network, and English word network\footnote{For the detailed description of data, see Table 2 in Ref.~\cite{Wu2017NeighborNeighbor}.}, resulting in $H_{\rm md}\approx 0.75$--$0.90$. In all networks analyzed, they found that the median-based, degree-dependent paradox holding probability $h_{\rm md}(k)$ is overall decreasing with $k$. These findings were to some extent explained by the network models preserving the empirical joint degree distribution $e_{kk'}$ while randomizing all other correlations. The model calculation was in good agreement with the empirical values of $H_{\rm md}$, but it overestimated (underestimated) the empirical $h_{\rm md}(k)$ for small $k$ (large $k$). A more complex model exploiting the joint degree distribution of three connected nodes explained better $h_{\rm md}(k)$ for the entire range of $k$. The joint degree distribution of three connected nodes was later used to define the correlation coefficient between degrees of two neighbors of a focal node, which was called transsortativity~\cite{Ngo2020Transsortative}.

Barone and Coscia~\cite{Barone2018Birds} analyzed a network of Italian firms, being either customers or suppliers or both ($N\approx 45$K and $L\approx 848$K), which were sampled from audited firms of Tuscany in 2007. They found that $\langle k\rangle_{\rm nn}(k)>k$ for nodes with $k<100$, composing around 92\% of nodes in the network.

Lee~\cite{Lee2026Friendshipparadox} reported that both the FP-net-link and FP-net-ego in Eqs.~\eqref{eq:FP_net_link_define} and~\eqref{eq:FP_net_ego_define} hold for the Zachary's karate club network ($N=34$)~\cite{Zachary1977Information} as well as for the American football network ($N=115$)~\cite{Girvan2002Community}. The author found that the value of $\rho_{kk_{\rm nn}}$ in Eq.~\eqref{eq:diff_ego_link_Lee} is negative (positive) for the Zachary's karate club network (American football network).

The FP also raises the question of how people feel about their own degrees. Yan et al.~\cite{Yan2022What} collected the responses from $N\approx 85.5$K Facebook users around the world in the summer of 2021 about how satisfied they were with their degrees on Facebook. 47\% (9\%) of the users preferred to have more (fewer) friends, while the rest of users were satisfied with their degrees. They found that $H=78.8\%$ of the surveyed users had fewer friends than the average degree of their friends, and that the degree-dependent paradox holding probability $h(k)$ decreased with $k$, finally vanishing for users with more than $3,500$ friends. The fraction of users who prefer more friends increased with the degree, even in the range where no FP exists. The authors concluded that the paradox itself was not enough to explain such preferences for high degrees.

\subsection{Generalized friendship paradoxes}\label{subsec:empirical_GFP}

The GFP has been empirically studied using network data with various nodal attributes whether they are topological or nontopological. Topological attributes of nodes are those derived only by the network topology. Grund~\cite{Grund2014Why} analyzed the same subnetwork of eight Marketville girls in Feld's original paper~\cite{Feld1991Why} to show that the GFP-net-ego in Eq.~\eqref{eq:GFP_net_ego_define} holds for the subnetwork when using closeness, betweenness, and eigenvector centralities~\cite{Newman2018Networks} as nodal attributes. He then analyzed friendship networks from 14 US middle schools to find that the GFP-net-ego held in every school for the degree as well as for the three mentioned centralities. It means that your friends are on average not only more popular but also more central in the network than you are.

Kong et al.~\cite{Kong2019How} analyzed the largest connected component of the DBLP computer science bibliography coauthorship network ($N\approx 1.5$M and $L\approx 7.2$M). To score each author, they modified the PageRank~\cite{Brin1998Anatomy} to define an academic rank, which is derived from the network topology alone. They first randomly sampled up to 100 authors for each degree $k$ and took the average of their academic ranks, i.e., $\langle x\rangle(k)$ in Eq.~\eqref{eq:avg_x_k} with $x$ denoting the academic rank. Then they obtained the average academic rank of coauthors of those authors, i.e., $\langle x\rangle_{\rm nn}(k)$ in Eq.~\eqref{eq:x_nn_k}, to find that $\langle x\rangle_{\rm nn}(k)>\langle x\rangle(k)$ for $k\lesssim 200$, confirming the degree-dependent GFP for most authors in the data.

Next, we review the results of the GFP using nontopological nodal attributes. By analyzing a two-month window of the 2009 Twitter data with $N\approx 3.4$M users and $L\approx 144.5$M links~\cite{Yang2011Patterns, Kwak2010What}, Hodas et al.~\cite{Hodas2013Friendship} showed that your friends are on average more active than you in terms of the number of tweets, and that your friends both post and receive URLs whose cascades are on average larger than yours. These observations were called activity paradox and virality paradox, respectively. The activity paradox held for $88\%$ of the users, while $86\%$ ($76\%$) of the users posted (received) less viral content than their friends did.

Eom and Jo~\cite{Eom2014Generalized}, who coined the term of GFP in 2014, analyzed two coauthorship networks mentioned in the previous subsection, namely, the PR corpus for 1893--2009 ($N\approx 243$K) and a snowball-sampled GS network ($N\approx 30$K). In addition to the number of coauthors, i.e., the degree, they considered the number of citations, the number of publications, and the average number of citations per publication for the PR network, and the number of citations for the GS network. For all considered attributes (including the degree), the paradox holding probabilities turned out to be $H_{\rm at}\approx 0.72$--$0.93$ [see Eq.~\eqref{eq:GFP_H_define}], and the GFP-net-link in Eq.~\eqref{eq:GFP_net_link_define} held in both PR and GS data. They found that the degree- and attribute-dependent paradox holding probability $h(k,x)$ in Eq.~\eqref{eq:GFP_hkx_define} decreases with $x$ for a given $k$ in all cases, while it increases with $k$ for a given $x$ only in the case when $x$ denotes the number of publications in the PR network. They attributed such an increasing behavior to the combined effect of the positive degree assortativity ($\rho_k\approx 0.47$) and the positive correlation between degrees and the numbers of publications ($\rho_{kx}\approx 0.79$).

In a similar line of research, Benevenuto et al.~\cite{Benevenuto2016Hindex} considered the h-index of the researcher~\cite{Hirsch2005Index} as the nodal attribute. They analyzed DBLP coauthorship networks derived from flagship conferences of ten Association of Computing Machinery Special Interest Groups (ACM SIGs) with $N$ being of the order of thousands, and estimated h-indexes of authors in those networks. Across all networks, 69--81\% of authors have h-index values lower than the average of their coauthors, which was called the h-index paradox. They attributed it to the positive correlation between the h-index and the degree, which was measured to be $\rho_{kx}\approx 0.36$.

Coletto et al.~\cite{Coletto2017Adult} analyzed the reblogging pattern of $N\approx 105$K adult-content users in Tumblr. Reblogging is the action by which users share the content of others. They found that $H_{\rm at} \approx 0.71$ in terms of the volume of the adult content reblogged by users, when restricting the comparison to friends who had been active at least once in the one-month observation window. They also reported that the Tumblr follower network rarely exhibits the majority illusion phenomenon~\cite{Lerman2016Majority} regarding the adult content. Thus, the perception bias was weak when only the degree is considered, whereas strong bias was observed when the volume of content shared by users was taken into account.

Bollen et al.~\cite{Bollen2017Happiness} proposed the happiness paradox, which was tested using a Twitter network with $N\approx 102$K users and $L\approx 2.4$M links~\cite{Bollen2011Happiness}. Among them, they analyzed $39,110$ users with at least $15$ friends and having nonzero subjective well-being (SWB). Each user's SWB was quantified via the sentiment analysis of up to $3,200$ tweets; the value of SWB ranges from $-1$ (most unhappy) to $1$ (most happy). The paradox holding probability using the SWB as the nodal attribute turned out to be $H_{\rm at}\approx 0.59$, which was compared to the baseline value of $\approx 0.50$ from the shuffled data, implying that the happiness paradox is a genuine effect rather than a statistical artifact. They also identified two subgroups of users, namely, happy group and unhappy group, to find that the unhappy group experienced the GFP more strongly than the happy group, while the FP in terms of the degree was stronger for the happy group.

Zhou et al.~\cite{Zhou2020Sentiment} studied the sentiment paradox stating that an individual's neighbors are on average more positive than the individual. Analyzing the LiveJournal network data ($N\approx 115$K users and $L\approx 246$K links)~\cite{Jin2017Emotions}, they defined the SWB as $(n_{\rm p} - n_{\rm n})/(n_{\rm p} + n_{\rm n})$ with $n_{\rm p}$ and $n_{\rm n}$ respectively denoting the numbers of positive and negative posts of a user. They reported that $H_{\rm at} \approx 0.55$, compared to the value of $\approx 0.50$ obtained by randomly reassigning the SWB values over the same network. They also tested the GFP at the triad and community levels, finding that the paradox persists at these coarser scales as well.

Luceri et al.~\cite{Luceri2025Susceptibility} studied the susceptibility paradox by analyzing Twitter data. They constructed two user-interaction networks, i.e., one for the 2020 US presidential election ($N\approx 50$K and $L\approx 117$K)~\cite{Chen2022Election2020} and the other for the COVID-19 pandemic ($N\approx 74$K and $L\approx 164$K)~\cite{Chen2020Tracking}. Each user's susceptibility is measured using how often the user shared the URLs exposed by their neighbors. Using this susceptibility as the nodal attribute, the authors found for both networks that the paradox holding probability $H_{\rm at}$ in Eq.~\eqref{eq:GFP_H_define} is larger than $0.5$, and that the GFP-net-link holds. However, the observed value of $H_{\rm at}=0.541$ in the interaction network for the election data is smaller than $0.672$ and $0.798$ obtained by randomizing the network while preserving the degree distribution and only the total number of links, respectively. The anti-GFP was found to hold when the nodal attribute is the number of URLs shared without prior exposure. They also measured degree- and attribute-dependent probabilities $h(k,x)$ in Eq.~\eqref{eq:GFP_hkx_define} to find that they show overall similar patterns to those for the coauthorship networks in Ref.~\cite{Eom2014Generalized}.

Surveys of self-perception revealed GFP-related patterns without measuring any network structure. Analyzing the data of 11 studies ($N=3,293$ in total), Deri et al.~\cite{Deri2017Home} found that most respondents rated their own social lives as poorer than those of others, in terms of the number of friends, the size of social networks, and the frequency of attending parties, among others. They argued (i) that the FP may not be sufficient to explain this result because the bias also appeared for items unrelated to the network structure such as the frequency of dining out, and (ii) that respondents compared themselves with the most socially active people among their neighbors rather than with an average person.

Finally, Chelmis and Gergin~\cite{Chelmis2023Recipe} analyzed a network of $N\approx 66$K recipes where links are formed when two recipes share the same ingredients. Each recipe $i$ is assigned a healthiness score $x_i$ measured by the United States Department of Agriculture; $x_i$ takes a value between $0$ (totally unhealthy) and $7$ (very healthy). They found that the FP holds for most nodes, and that the GFP holds for nodes with $x_i\leq 3$. It implies that relatively unhealthy recipes have on average healthier neighbors than themselves, which they called a healthiness paradox.

\subsection{Friendship indexes and neighborhood gaps}\label{subsec:empirical_FI_gap}

Pal et al.~\cite{Pal2019Study} computed the arithmetic, geometric, and harmonic means of the FI in Eqs.~\eqref{eq:arith_mean_FI_define}--\eqref{eq:harmonic_mean_FI_define} for 13 real networks spanning five categories; social networking services (Hamsterster, Brightkite, Douban, Gowalla, Hyves), human social networks (Jazz musicians, Zachary's karate club), human contact networks (Hypertext, Infectious), computer networks (Route Views, Internet topology), and infrastructure networks (Power grid, Euroroad)\footnote{For the detailed description of data, see Table 1 in Ref.~\cite{Pal2019Study}.}. They reported the values of $\langle I\rangle$ ranging from $\approx 1.44$ for Euroroad to $\approx 786$ for Hyves, while those of $\langle \log I\rangle$ range from $\approx 0.23$ for Euroroad to $\approx 4.32$ for Internet topology. They also found that the local assortativity defined for a node~\cite{Thedchanamoorthy2014Node} shows a strong, negative correlation with the value of $|\log I_i|$ in most networks considered; nodes in assortative neighborhoods have $I_i \approx 1$, whereas those in disassortative neighborhoods have extreme values of $I_i$. These effects of local degree correlation on the strength of the FP can complement the explanation based on the network-level degree assortativity.

Sidorov et al.~\cite{Sidorov2021Friendship} measured the distribution of the FI, denoted by $P(I)$, to find heavy tails in $P(I)$ for four undirected networks, i.e., Amazon co-purchase network ($N \approx 335$K), mobile phone user network ($N \approx 37$K), GitHub mutual follower network ($N \approx 38$K), and Facebook artists' network ($N \approx 51$K), as well as for two directed networks, i.e., literature citations ($N \approx 450$K) and web hyperlinks ($N \approx 875$K). By assuming the functional form of $P(I)\propto I^{-\gamma}$, they estimated the power-law exponent $\gamma$ from $1.1$ (GitHub) to $2.7$ (mobile phone users). The fraction of nodes satisfying $I_i > 1$, corresponding to $H$, ranges from $75\%$ for web hyperlinks to $99\%$ for GitHub. Such heavy-tailed FI distributions imply that a considerable fraction of nodes experience an extensive disparity with neighbors having degrees very different from their own degrees.

Jackson~\cite{Jackson2019Friendship} computed the distribution of the FI, $P(I)$, for a rural Indian village network of 135 households~\cite{Banerjee2013Diffusion}, where nodes are households and links are formed between nodes when they exchange favors such as rice. The observed $P(I)$ shows a peak at $I\approx 1$ but also a right-skewed tail up to $I\approx 4.5$, leading to the average value of $\langle I\rangle\approx 1.43$.

The FI was also employed to study the temporal dynamics of the FP in the smartphone communication data. Wang et al.~\cite{Wang2025Temporal} analyzed the NetHealth data, taking the 119 days of the fall semester of 2015 at the University of Notre Dame ($N=692$ students and $L=9,732$ links between them). They found that the FI values stabilized faster than the individual degrees and the spread of their friends' degrees did. In addition, the same authors~\cite{Wang2021Using} used the same data (but with different observation period) to confirm the GFP in terms of daily physical activity status, the number of steps walked, activity duration, and the amount of burnt calories.

Lattanzi and Singer~\cite{Lattanzi2015Power} focused on the ratio of the average degree of neighbors and the average degree in a set of $n$ randomly chosen nodes, which is a slightly different version of the FI:
\begin{align}
    I_n\equiv \frac{\langle k_j\rangle_n}{\langle k_i\rangle_n},
\end{align}
where for each sampled node $i$, one random neighbor $j$ of $i$'s is chosen and included for the calculation of $\langle k_j\rangle_n$. They studied the effect of sample size $n$ on $I_n$ for eight empirical network datasets (Orkut, LiveJournal, Wikipedia, YouTube, DBLP, Slashdot, Enron, and Epinions)\footnote{For the detailed description of data, see Table 1 and related references in Ref.~\cite{Lattanzi2015Power}.} to find that $I_n$ tends to decrease as $n$ grows, while staying well above $1$. They concluded that the neighbor of a random sample tends to have a higher degree than the sample itself does in all datasets, with the ratio averaged over samples ranging from $2.5$ (DBLP) to $150$ (YouTube). An asymptotic version of this ratio was studied for a large family of random graphs.

Instead of the FI, Cantwell et al.~\cite{Cantwell2021Friendship} focused on the neighborhood gap. They measured the average and variance of neighborhood gaps, $\Delta k_i$ in Eq.~\eqref{eq:deltak_i}, using 32 empirical social networks\footnote{For the detailed description of data, see the references in Ref.~\cite{Cantwell2021Friendship}.}. The average and variance of neighborhood gaps were also computed analytically for the model networks having the same degree distributions with or without the same values of the degree assortativity as the empirical data. It was found that the empirical and model-based neighborhood gaps were in good agreement even without considering the degree assortativity, while consideration of the degree assortativity improved the agreement between them.

Evtushenko and Kleinberg~\cite{Evtushenko2023Nodebased} analyzed the Facebook100 dataset containing anonymized friendship networks at 100 US institutions~\cite{Traud2012Social}, with the average network size of $N\approx 12.8$K users. As for the nodal attribute, they considered the fraction of each user's friends who have the same reported gender as the user. The authors found that for most of 100 networks the GFP-net-link and the GFP-net-ego either both hold or both fail to hold, whereas 9 out of 100 networks fell into the case where the GFP-net-link holds but the GFP-net-ego does not (see Sec.~\ref{subsec:gap}).

\subsection{Results in directed, weighted, and signed networks}\label{subsec:empirical_directed_weighted_signed}

Here we will overview the literature on results in directed networks, weighted networks, and signed networks one by one. Twitter has served as the primary empirical platform for testing the directed FP as formulated in Sec.~\ref{subsec:direct_weight}. Using the Twitter data with $N\approx 5.8$M users and $L\approx 194$M directed links~\cite{Yang2011Patterns}, Hodas et al.~\cite{Hodas2013Friendship} provided the first empirical evidence for the FP and GFP on the directed network. For each user $i$, they essentially computed the directed versions of the friendship index in Eq.~\eqref{eq:FI_define}:
\begin{align}
    I^{\alpha|\beta}_i \equiv \frac{k^{\alpha|\beta}_{i,\rm nn}}{k^{\alpha}_i}
\end{align}
for $\alpha,\beta\in\{\rm in, out\}$ [see also  Eq.~\eqref{eq:kinn_directed_define}]. All distributions of those friendship indexes, except for $I_i^{\rm out|in}$, show peaks at around $10$--$20$ and are right-skewed. For all combinations of $\alpha,\beta\in\{\rm in,out\}$, they confirmed the degree-dependent FPs using Eq.~\eqref{eq:k_nn_k_direct} for the ranges of $k^{\rm in}, k^{\rm out} <10^3$, and all the values of $H^{\alpha|\beta}$ in Eq.~\eqref{eq:FP_direct_H_define} were estimated to be at least $0.98$.

Kooti et al.~\cite{Kooti2014Network} refined the above analysis on the same Twitter data as well as on a Digg follow network with $N\approx 11.9$K users and $L\approx 1.3$M links. A majority of users experienced the median-based FP in terms of both followers and followees. They applied both mean-based and median-based comparisons to activity, content diversity, and URL virality, all of which exhibited the GFP. However, a shuffle test randomizing nodal attributes revealed that the median-based FPs and GFPs mostly disappeared, while the mean-based ones persisted except for the virality of content received on Digg. It implies that the median-based FPs and GFPs relied not only on the degree or attribute heterogeneities but also on the correlation between a node's degree and its attribute and the correlation between the attributes of neighboring nodes.

Momeni and Rabbat~\cite{Momeni2016Qualities} studied the mean-based and median-based GFPs in directed networks using the Twitter data~\cite{Yang2011Patterns, Kwak2010What}. They considered two topological measures (in-degree and out-degree), two activity measures (the number of original tweets and the number of tweets including retweets), and four influence measures based on the number of retweeted tweets. For topological measures, they computed the values of $H^{\alpha|\beta}$ in Eq.~\eqref{eq:FP_direct_H_define} for $\alpha,\beta\in\{\rm in,out\}$ using both mean-based and median-based definitions. For activity and influence measures, they estimated values of $H_{\rm at}$ in Eq.~\eqref{eq:GFP_H_define} but defined on the directed network, again using both mean-based and median-based definitions. In almost all cases, more nodes experienced the FP or GFP when they were compared to their followees than when compared to their followers. The authors called it a hierarchy of attachment, meaning a tendency to follow those higher degree or attribute than themselves. They also measured the degree-dependent paradox holding probabilities in Eq.~\eqref{eq:FP_hk_direct_define} and the directed version of attribute-dependent paradox holding probabilities in Eq.~\eqref{eq:GFP_hk_hx_define}. Most curves increase or stay flat before eventually approaching zero, and the curve for the in-degree was noted to agree with $h(k)$ obtained for model networks in Ref.~\cite{Jo2014Generalized}.

Alipourfard et al.~\cite{Alipourfard2020Friendship} collected a Twitter subgraph in 2014 with $N \approx 5.6$K seed users of average degree around $124$ and their friends, ending up with $N \approx 600$K users in total. They measured the degree-dependent paradox holding probabilities $h_\beta(k^\alpha)$ in Eq.~\eqref{eq:FP_hk_direct_define} for $\alpha,\beta\in\{\rm in,out\}$, all of which overall decreased with $k^{\rm in}$ or $k^{\rm out}$.

Moving beyond Twitter, Iyer~\cite{Iyer2018Friendship} randomly chose $100$K users of an online knowledge depository, Quora, who have visited Quora in the four weeks preceding June 1, 2016 and have at least one follower and one followee. For each user, the author computed average and median in-degrees and out-degrees of in-neighbors and out-neighbors to confirm all the four median-based FPs defined in the directed network.

Serafimov et al.~\cite{Serafimov2019Friendship} paid attention to the Instagram data; they analyzed the directed network with $N\approx 45$K users and $L\approx 678$K links between them by employing nine nodal attributes, i.e., in-degree, out-degree, numbers of received/given likes/comments, and numbers of media posts, total hashtags, and unique hashtags. They computed the network-level paradox holding probabilities $H^{\alpha|\beta}$ in Eq.~\eqref{eq:FP_direct_H_define} for the in-degree and out-degree to be $0.74$--$0.92$. The directed versions of $H_{\rm at}$ in Eq.~\eqref{eq:GFP_H_define} for the other attributes were measured to be $0.59$--$0.86$. In addition, the median-based network-level paradox holding probability $H_{\rm md}$ turns out to be smaller than $H$ for every attribute, yet those values are greater than $0.5$, except for the numbers of total/unique hashtags of followees. These findings are reminiscent of the followee-follower asymmetry reported in Ref.~\cite{Momeni2016Qualities}.

The directed FP has also been tested outside online platforms. Mansueli et al.~\cite{Mansueli2023Friendship} constructed invitation networks of academic thesis defense committees from two open-access repositories, namely, the Brazilian Capes Catalog (Capes) and the French STAR Deposit (STAR). Nodes in the network include advisors of the theses defended (inviters) and invited committee members (invitees); some members play both roles of the inviter and the invitee, leading to $N\approx 117$K (Capes) and $\approx 78$K (STAR). Each directed link runs from the inviter to the invitee. They tested two directed paradoxes in the form of Eq.~\eqref{eq:kinn_directed_define}: (i) the invitee paradox states that invitees are invited on average fewer times than those who invited them ($k_{i,\rm nn}^{\rm in|in}>k_i^{\rm in}$), and (ii) the inviter paradox states that inviters invite on average fewer times than those whom they invited ($k_{i,\rm nn}^{\rm out|out}>k_i^{\rm out}$). They found that only the invitee paradox holds. Consistently, the network-level paradox holding probability for the invitee paradox is $H^{\rm in|in} \approx 0.77$ (Capes) and $\approx 0.58$ (STAR), while those for the inviter paradox satisfy $H^{\rm out|out} < 1/2$ for both datasets.

Next, one can understand the paradoxes in a more nuanced way by taking link weights into account. Bagrow et al.~\cite{Bagrow2017Which} sampled $3,300$ Twitter users and around $90$K mobile phone users in a European country as well as their egocentric networks for both datasets. In each egocentric network, alters are ranked according to their interaction or communication volumes with the ego. Then the ego's out-degree is compared to the mean or median out-degrees of their $n$th ranked alters. They found that the most contacted alters to egos tend to have less out-degrees than the less contacted alters, and that for the mobile phone data, the egos had on average higher out-degrees than their most contacted alters, implying the opposite of the FP. Yet they observed that the network-level paradox holding probabilities are $H\approx 99.7\%$ for Twitter and $\approx 76.8\%$ for mobile phone data, which are driven not by the so-called best friends but by more distant friends.

Finally, the paradoxes were studied in the empirical signed network with both positive and negative links, denoting friendship and enmity, respectively. Ghasemian and Christakis~\cite{Ghasemian2023Enmity} analyzed the social network of $N\approx 25$K individuals covering 176 villages in rural Honduras. They reported that the average number of friends is $6.89$, while the average number of friends of those friends is $8.40$, implying the network-level FP\footnote{It is not clear whether it is FP-net-link or FP-net-ego from the text ``these friends have an average of 8.40 {(SD = 2.52)} friends'' in Ref.~\cite{Ghasemian2023Enmity}.}. It was found that the average number of enemies is $1.26$, while the average number of enemies of those enemies is $3.40$, implying the network-level enmity paradox. The generalized enmity paradox was also confirmed with respect to the wealth as the nodal attribute, i.e., your enemies are on average richer than you are, though it was weaker than its friendship counterpart. They tested a multistep FP, also discussed in Ref.~\cite{Kramer2016Multistep}, with the step number $l=1,\ldots,6$, where $l=1$ corresponds to the original definition of the FP. The larger $l$ resulted in bigger neighborhood gaps between the average degree of nodes and the average degree of multistep neighbors whether the neighbors denote friends or enemies. It is because sampling biases due to the degree heterogeneity accumulate exponentially as $l$ increases, so that high-degree nodes are increasingly overrepresented among the multistep neighbors.

\section{Applications}\label{sec:applications}

The original FP and its variants including GFPs have found a number of applications across various domains from early detection of outbreaks and influence maximization to network sampling and perception biases. We remind that the probability of sampling a node with degree $k$ by following a randomly chosen link is proportional to $kP(k)$ rather than $P(k)$~\cite{Newman2003Egocentered, Newman2003Why, Newman2007Course}, leading to the overrepresentation (underrepresentation) of high-degree nodes (low-degree nodes) in the sample. Therefore, looking at one's friends may give a biased but far richer view of the network than looking at oneself. The FP-based sampling bias can be characterized, corrected, and/or employed for various purposes~\cite{Galesic2021Human, Feld2020Egonets}. We review the literature on the aforementioned topics one by one, though they are not necessarily mutually exclusive.

\subsection{Early detection and mitigation of outbreaks}\label{subsec:outbreak}

The FP has been used as a principle behind the sensor method~\cite{Christakis2010Social, Garcia-Herranz2014Using, Kryvasheyeu2015Performance, Sun2014Efficient, Mathias2026Friends}. The sensor method exploits the fact that a random person's friends are on average more central in the network than that person, as demonstrated in Refs.~\cite{Grund2014Why, Higham2019Centralityfriendship, Higham2026Generalized, Hazra2026Generalized}. Central nodes are more likely exposed to infectious disease or information spreading in networks than non-central, peripheral nodes. Therefore, monitoring a sample of neighbors of random nodes (friend or sensor group) is expected to detect outbreaks earlier than monitoring a sample of random nodes (random or control group). Results using the sensor and control groups are often compared to those using the highest-degree nodes (hub group), but only when the global information on the network is accessible. The sensor method can shed light on how to mitigate outbreaks effectively too, e.g., by helping to identify nodes who are more susceptible to the disease or misinformation.

Christakis and Fowler~\cite{Christakis2010Social} provided the first real-world test of the sensor method by tracking a flu outbreak among 744 Harvard undergraduates in the fall of 2009. Each of $425$ students in the friend group was named as a friend by at least one of $319$ students in the control group. Compared to the control group, students in the friend group had significantly higher values of in-degree and betweenness centrality, and they caught the flu earlier. Accordingly, both the cumulative curve and the peak time of the daily number of flu incidents for the friend group were shifted forward in time by $13.9$ days, compared to the control group. These findings indicate the early detection of the outbreak by the sensor method.

Garcia-Herranz et al.~\cite{Garcia-Herranz2014Using} applied the same sensor method to hashtag usage patterns in Twitter by analyzing the Twitter following network in 2009 ($N=40$M users and $L=1.5$B directed links)~\cite{Kwak2010What}. $50$K random users were sampled to make the control group, and those users' random followees comprised the sensor group of the same size. For each of 32 most popular hashtags posted by the control group, the average timing when users in the sensor group posted the hashtag for the first time was compared to that in the control group. A lead time is defined as the average timing for the sensor group minus that for the control group; the negative value of the lead time is expected due to the FP-based bias. It turned out that the lead times are negative for most hashtags. Users in the sensor group posted the popular hashtags on average $7.1$ days prior to those in the control group.

Kryvasheyeu et al.~\cite{Kryvasheyeu2015Performance} also employed the sensor method to disaster monitoring by analyzing over $52$M posts in Twitter from nearly $14$M users during Hurricane Sandy in 2012. The control group was randomly sampled from the users whose locations were identified, and the sensor group was sampled as the followees of users in the control group. They reported that sensor groups exhibited an advantage of disaster awareness in terms of the lead time up to $-26$ hours over control groups. The extent of the advantage varied with the sample size and the users' locations relative to the storm. The lead time was the longest when the control group was drawn from outside the hurricane zone while the sensor group was drawn from within it. This result serves as an example of the interplay between network centrality and geographic factor.

Sun et al.~\cite{Sun2014Efficient} demonstrated the sensor method for a city-wide physical contact network reconstructed from one week of smart-card transit records of Singaporean bus users. The data comprises $1$B pairwise contacts between $3$M users. The control group is a 1\% random sample of the users, and the friend group consists of one randomly chosen neighbor of each member of the control group. The friend group has nearly twice the average degree of the control group ($442$ versus $239$), and it shows similar biases in travel frequency, $k$-shell index, and encounter entropy\footnote{Here $k$-shell index is the largest $k$ such that the node remains when nodes with degree less than $k$ are repeatedly removed, and the encounter entropy is the Shannon entropy of the hourly distribution of an individual's encounters.}. In addition, the authors performed numerical simulations of the susceptible-exposed-infected (SEI) epidemic process on the empirical network. They found that the friend group experienced infection ahead of the control group. It was also shown that monitoring the 0.01\% highest-degree users (hub group) resulted in earlier and more consistent warning than monitoring the friend group. Therefore, when the contact network is fully observable, direct monitoring of highest-degree nodes dominates monitoring the friend group.

Nettasinghe et al.~\cite{Nettasinghe2020Diffusion} explored the effects of the FP on contagion processes in which nodes are affected by their neighbors' neighbors (thus called monophily~\cite{Altenburger2018Monophily}) rather than their immediate, nearest neighbors. By analyzing the mean-field dynamics of susceptible-infected-susceptible (SIS) epidemic processes in networks, they found that the epidemic threshold for the monophilic contagion is no higher than that for the nearest-neighbor contagion. It implies that the spreading is more prevalent in the monophilic contagion when all other conditions are equal. They also found that among networks sharing the same degree distribution the disassortative networks lowered the epidemic threshold for the monophilic contagion because the degree disassortativity amplifies the FP.

The FP was also leveraged to mitigate outbreaks or epidemic spreading by means of immunization~\cite{Pastor-Satorras2002Immunization, Pastor-Satorras2015Epidemic} and contact tracing~\cite{Kiss2005Disease, Kiss2006Infectious, Bassolas2022Optimizing}. Epidemic spreading and immunization in networks have been translated into the problems of percolation, resilience, and robustness~\cite{Albert2000Error, Callaway2000Network, Cohen2000Resilience, Holme2002Attack, Crucitti2004Error, Altarelli2014Containing, Liu2020Efficient, Li2022Sitebond}. Several immunization strategies have been proposed in the literature, among which the acquaintance immunization~\cite{Cohen2003Efficient} utilizes the FP-based sampling bias as it targets random neighbors of random nodes (sensor group). As a baseline, the random immunization targeting randomly chosen nodes (control group) has been considered. Finally, these strategies are often compared to the hub immunization by which the highest-degree nodes (hub group) are targeted. All groups are designed to have equal or similar sizes both for fair comparison and because of the limited budget for interventions such as immunization.

Cohen et al.~\cite{Cohen2003Efficient} derived an analytical equation for the critical fraction of nodes immunized to halt the outbreak, denoted by $f_{\rm c}$, when using the acquaintance immunization. When the power-law degree distribution $P(k)\propto k^{-\gamma}$ is considered for networks of $N=10^6$, the value of $f_{\rm c}$ turns out to be around $0.25$ for $2<\gamma<3$. In contrast, when randomly chosen nodes are immunized for the same networks, $f_{\rm c}$ approaches $1$ for the same range of $\gamma$, essentially meaning that the entire population should be immunized to halt the outbreak. More detailed analysis and test results are found in Ref.~\cite{Madar2004Immunization}.

Manzo and van de Rijt~\cite{Manzo2020Halting} tested the effectiveness of various immunization methods. Everyday the same number $n$ of individuals are sampled among non-immunized people for the immunization, that is, $n$ randomly chosen individuals (random immunization), $n$ random neighbors of $n$ random individuals (acquaintance immunization), and $n$ highest-degree individuals (hub immunization). They simulated the susceptible-exposed-infectious-recovered (SEIR) epidemic process under each of three immunization methods on networks derived from the COMES-F close-range contact survey in France ($N=2,029$). Because the survey does not indicate how a respondent's neighbors are interconnected, the empirical value of clustering coefficient $C$ is unknown. Using the empirical degree distribution, they generated networks whose $C$ can vary across the full range of $0.01$--$0.57$ for the given degree distribution. For the case with $C=0.01$ and $n=10$, the random, acquaintance, and hub immunization methods respectively resulted in the infection peaks of $174$, $62$, and $16$, compared to the peak of $437$ without immunization.

Yang et al.~\cite{Yang2019Efficient} employed five immunization methods, namely, vaccinating randomly chosen individuals (random), one contact named at random by each randomly chosen individual (nomination or acquaintance), randomly interviewed high-degree individuals (high degree), the top $10\%$ of individuals by degree (highest degree), and the top $10\%$ of individuals by betweenness centrality (central). They simulated the susceptible-infected-recovered (SIR) epidemic process on the fully observed topology of $75$ empirical Indian village networks. The random vaccination reduced cumulative incidence by $32.3\%$ relative to no intervention, and the nomination vaccination reduced it by an additional $10.7\%$ without the information on the entire network. In contrast, the highest-degree and central vaccination, requiring access to the entire network, reduced the cumulative incidence by about two-thirds relative to the random vaccination.

Rosenblatt et al.~\cite{Rosenblatt2020Immunization} tested immunization strategies in networks with missing data. They generated synthetic networks using the degree distribution of the Colorado Springs network of at-risk individuals for HCV and HIV transmission~\cite{Klovdahl1994Social} but with different levels of transitivity. For each generated network, a fraction of randomly chosen nodes are considered missing. Then target nodes are identified and immunized using each immunization strategy. The susceptible-infected-recovered (SIR) epidemic process is then simulated on each immunized network to give the outbreak size, which is used to test the effectiveness of immunization methods. When the fraction of missing nodes is small, the best strategies are degree immunization and betweenness immunization, both requiring the global information on the network. However, once more than half of nodes are missing, the variants of the acquaintance immunization appear to be more effective than others.

Coviello et al.~\cite{Coviello2019predicting} also compared immunization methods by simulating the susceptible-infected (SI) epidemic process on the temporal network. From the online review site Yelp, they derived a static friendship network of $N\approx 169$K users and a temporal, encounter network of the same users. Links in the encounter network represent the physical encounters of users; the link between two users appears only when they post reviews to the same business on the same day. Three immunization methods are considered: (i) The random immunization targets randomly chosen nodes; (ii) The acquaintance immunization targets random friends of random nodes in the friendship network; (iii) The encounter immunization targets future encounters of random nodes in the encounter network. Here the future encounter of a node is not necessarily the node's neighbor in the friendship network. By the simulations on the encounter network, they found that the encounter immunization outperforms the acquaintance immunization, yet both are more effective than the random immunization as expected.

Kojaku et al.~\cite{Kojaku2021Effectiveness} demonstrated an even more effective strategy than the acquaintance immunization, namely, the backward contact tracing. The backward contact tracing samples individuals (or parents in the transmission tree) who infected the confirmed cases, rather than the confirmed cases' children nodes in the tree. The probability of sampling such parent nodes is proportional to $k(k-1)P(k)$ or $\propto k^2P(k)$ for a large $k$, showing a stronger bias than the FP. They performed simulations of susceptible-exposed-infectious-recovered (SEIR) epidemic process on the Barab\'asi--Albert (BA) scale-free network with $N=250$K and on the empirical Copenhagen Networks Study contacts with $N=567$. The backward contact tracing indeed reduced the epidemic peak by more than $70\%$ relative to the control group on the BA network, and cut the peak by about $50\%$ compared to about $15\%$ for the control group on the empirical network. 

We finally remark that the highest-degree nodes or hubs are not necessarily optimal to be sampled for observing dynamics on networks. For example, optimal nodes or sentinels for observing overall dynamics of networked systems tend to avoid central nodes such as hubs~\cite{MacLaren2025Observing}, and hubs are found to suppress synchronization of coupled oscillators~\cite{Motter2005Network}. These findings imply that hubs may play an adversarial role depending on the nature of dynamics, hence they are not always best to represent or sample the systems.

\subsection{Seeding for information dissemination}\label{subsec:seeding}

Another domain of network interventions using the FP can be found in the context of seeding for information dissemination or influence maximization in networks~\cite{Kempe2003Maximizing, Kitsak2010Identification, Valente2012Network, Chen2012Identifying, Pei2014Searching, Morone2015Influence, Pei2020Influencer}. The goal of the influence maximization is to identify a small set of nodes or seeds in a given network, who pass the information to their neighbors along the network, to maximize the number of informed nodes. More central nodes have been considered being more influential. However, calculation of network centralities of nodes requires the information on the entire network. When the entire network is not accessible, the FP-based seeding strategies are particularly useful as they do not require the global information on the network.

Han et al.~\cite{Han2014Your} devised a protocol to identify influential nodes in a mobile social network using a random-walk sampling. Probe messages are randomly generated in mobile devices and transmitted to other random devices during contacts, where only a small number of transmissions is allowed for each message. Then the most visited devices by the messages are chosen as being influential nodes. They simulated the information dissemination on the Dartmouth WiFi contact trace of $4,522$ devices, where two devices are considered being in contact whenever they are associated with the same access point. Using a seed set of $50$ mobile devices identified by their protocol, they reached $51\%$, $37\%$, and $14\%$ more devices compared to the case with a seed set of random nodes when the information transmission probabilities are $0.005$, $0.01$, and $0.05$, respectively.

Ben~Sliman and Kohli~\cite{BenSliman2018Asymmetric} tested three seeding methods in the directed network; the seed set consisted of either the followees of random nodes (followee group), the followers of random nodes (follower group), or random nodes themselves (random group). The average numbers of followers per seed are compared using the Twitter data ($N\approx 28.5$M)~\cite{Rossi2015Network}. The average numbers of followers for the followee, follower, and random groups were found to be around $1,600$, $1,400$, and $42$, respectively. It implies that seeding the followees of random nodes can reach the largest number of users.

Feng et al.~\cite{Feng2022Neighborhood} studied how the FP-based seeding can enhance influence maximization when the independent cascade model~\cite{Kempe2003Maximizing} is used for the diffusion process starting from a small subset of seed nodes in networks. They first proved a set-level FP; for any connected network, when each node is chosen with a vanishingly small probability, the average degree of those chosen nodes is no larger than that of their neighbors. Therefore, by seeding to neighbors of random nodes one can reach influential nodes more often than seeding to random nodes. They performed experiments on four real-world networks derived from Stanford Network Analysis Project (SNAP). For each network, $100$ random users are given an offer under the limited budget and then included to the set $S$ only when they accept the offer. Denoting the neighbors of $S$ by $\Lambda(S)$, they found that seeding using both $S$ and $\Lambda(S)$ results in the larger influence than seeding using only $S$.

McMillan and Schaefer~\cite{McMillan2021Comparing} used empirically grounded simulations to test whether FP-based intervention strategies could lower the drinking level of inoculation program participants and non-participants. Using friendship nominations and drinking reports from $N=4,819$ students in $28$ US high school cohorts, they fitted a social network dynamics model~\cite{Snijders2001Statistical} in which students' drinking and friendship can coevolve. Target students were selected either by network centralities (i.e., in-degree and betweenness) or by friend-bringing strategies (i.e., each randomly chosen student brings one, up to three, or all of their friends). All of these five strategies led to the lower drinking levels compared to the case with no interventions. While the strategies targeting high in-degree and betweenness students lowered the drinking level compared to the random targeting, all of three friend-bringing strategies did not show any significant improvement. It demonstrates that the FP-based strategies do not always improve the outcome.

Unlike the mentioned simulation studies, the following three works tested the effectiveness of FP-based intervention methods on real people. Banerjee et al.~\cite{Banerjee2019Using} seeded information on the immunization for children to targeted individuals in each of 521 villages of Haryana, India. For some villages, the targets were heads of randomly chosen households (random group), while individuals who can spread information to many villagers were nominated and targeted in other villages (gossip group). They found that $21\%$ more children were vaccinated in the gossip group than in the random group. 

Alexander et al.~\cite{Alexander2022Algorithms} also performed a field experiment to see the effectiveness of FP-based seeding algorithms. To promote iron-fortified salt for women, they seeded information to $20\%$ of women in each of $50$ buildings in Mumbai, India ($N=2,491$ in total). They targeted (i) random friends of random women (friend group) and (ii) pairs of random women and their friends (pair group), which were compared to targeting randomly chosen women (random group) and the case with no targeting. $21$--$22\%$ of women in friend and pair groups purchased the iron-fortified salt, whereas $13\%$ in the random group and $7\%$ for the case with no targeting did. 

Lastly, Airoldi and Christakis~\cite{Airoldi2024Induction} ran a similar field experiment on health intervention across $176$ isolated villages in Honduras ($N=24,702$ in total) using $117$ outcome measures. Comparing the friendship-nomination targeting to the random targeting, they found that the FP-based targeting was more efficient than the random targeting for $34$ of $117$ outcomes.

Finally, Lera and Leng~\cite{Lera2026Pairwise} discussed that the FP-based seeding strategy developed for pairwise networks may not be optimal in higher-order systems.

\subsection{Network sampling and polling}\label{subsec:sampling}

The FP has also been used as a sampling tool in networks. Network sampling matters when access to the entire network is limited due to incomplete or missing data on nodes and links or when network data are too large to download in full~\cite{Stumpf2005Sampling, Stumpf2005Subnets, Lee2006Statistical, Leskovec2006Sampling, Smith2013Structural, Smith2017Network, Torok2016What}. When the sample size is fixed due to the limited budget (costs for API or survey), one can reach more nodes by sampling high-degree nodes than average or low-degree nodes. Here the FP-based sampling bias is useful as it overrepresents high-degree nodes, but it should be corrected to give unbiased estimates for the entire network. Precisely, the FP-based sampling bias has been (i) exploited to efficiently sample high-degree nodes, allowing for better resolution of heavy tails in degree distributions, (ii) corrected to recover an unbiased estimate from a biased sample, and (iii) used to devise more efficient polling methods than the classical one, all assuming the limited access to the entire networks.

Maiya and Berger-Wolf~\cite{Maiya2011Benefits} related the network crawling to the FP. They compared seven network-crawling strategies by applying them to 12 social and technological networks. The crawling is an iterative process that starts with a single node (or seed) in the sample set $S$ and repeatedly adds one of neighbors of nodes in $S$ to $S$. On the top of existing strategies such as the breadth-first search and depth-first search, the authors devised the sample edge count (SEC) method that chooses a node having the largest number of links within $S$ for the addition. They found that the SEC outperforms other strategies for sampling high-degree nodes, yet it is one of the worst for recovering the entire degree distributions. For the task of sampling high-degree nodes, the SEC was compared to the FP-based acquaintance sampling method~\cite{Christakis2010Social}, by which random neighbors of random nodes are sampled; the SEC needed a smaller sample size than the acquaintance sampling for every network tested. 

Kurant et al.~\cite{Kurant2011Unbiased} carried the FP-based bias into the crawling of networks too large to download in full. Among many crawling methods, they focused on the breadth-first search (BFS) that induces bias toward high-degree nodes just as the FP does. The degree distribution of nodes sampled by the BFS, denoted by $q_k$, is asymptotically proportional to $kP(k)$ when the fraction of sampled nodes is small. By inverting the relation between $q_k$ and $P(k)$, the true degree distribution $P(k)$ can be estimated as $\propto q_k/k$. This corrected estimation was applied to Facebook and Orkut networks whose full topologies are unknown. In the case of Facebook, they ran $28$ small BFS crawls that together covered less than $1\%$ of the network. The corrected estimate of the average degree using the sampled nodes is $85.4$, which is comparable to the true value of $94.1$.

Eom and Jo~\cite{Eom2015Tailscope} exploited the FP to estimate heavy tails of degree distributions in large-scale networks. They also used the relation $P(k)\propto q_k/k$, where $q_k$ is obtained from nodes sampled by following randomly chosen links. In the case of scale-free networks with $P(k)\propto k^{-\alpha}$ ($\alpha>2$), they argued that the sampling resolution, defined as the maximum degree in the sample, is $n^{1/[(\alpha-2)(\alpha-1)]}$ times greater than that of uniform sampling, where $n$ is the number of sampled nodes. Therefore, sampling nodes by following random links results in the more resolved tail than a uniform sample of the same size does, which they called the tail-scope method. Across eight social and technological networks with up to $N\approx 42$M\footnote{For the detailed description of data, see the Methods section of Ref.~\cite{Eom2015Tailscope}.}, sampling $n=1,000$ nodes reproduced the heavy tail much better than uniform sampling did. However, the tail-scope method was worse than the uniform sampling to reproduce $P(k)$ for small degrees. Combining the uniform and tail-scope sampling recovered the true degree distribution for the entire range of degrees. 

Pires et al.~\cite{Pires2017Friendship} applied the FP to identify important species in ecological networks. They analyzed six plant-pollinator networks, each with more than $50$ species and with links weighted by the frequency of interaction between a plant and a pollinator. Firstly, $n$ plants or $n$ pollinators are chosen at random to make a random group. Then three most frequently interacting neighbors of each species in the random group are collected, among which $n$ most frequently appeared species are chosen to make a partner group. The authors ran the sampling process $100$ times for each network with $n=2$, $4$, and $8$, respectively. As predicted by the FP, species in the partner group has a tendency to have higher degree and betweenness than those in the random group. They also confirmed by dynamical model simulations that removing four species in the partner group from the network led to the extinction of more species than removing four species in the random group. The removal of the partner group disturbed the network structure more than that of the random group did.

Using the FP, Newman~\cite{Newman2003Egocentered} corrected an estimate of the number of friends of friends a person has, originally proposed by Bernard et al.~\cite{Bernard2001Estimating}. The original estimation assumes each friend as an average person. If a person has $\langle k \rangle$ friends, each of those friends adds $\langle k \rangle-1$ contacts beyond the tie back to that person, leading to the estimate of $\langle k \rangle(\langle k \rangle-1)$ friends of friends. However, the average degree of a friend is not $\langle k\rangle$ but $\langle k\rangle_{\rm link}=\langle k^2\rangle/\langle k\rangle$ in Eq.~\eqref{eq:k_link_result}. Therefore, the number of friends of friends of a person should be $\langle k \rangle(\langle k \rangle_{\rm link}-1)=\langle k^2 \rangle-\langle k \rangle$. This result exceeds the original estimate exactly by the variance of degrees, $\sigma_k^2$. Beyond the degree heterogeneity as a correction, the author also considered the effects of triangles and squares in networks in terms of the clustering coefficient $C$~\cite{Watts1998Collective} and mutuality $M$, respectively. $C$ measures the fraction of links between one's friends, while $M$ does the average number of paths of length 2 to one's second neighbor. The estimates corrected by these two factors indeed gave the most accurate results in the cases of two coauthorship networks in mathematics ($N\approx 253$K)~\cite{Newman2001Structure} and biomedicine ($N\approx 1.5$M)~\cite{Grossman1995Portion}.

Nettasinghe and Krishnamurthy~\cite{Nettasinghe2019What} applied the FP to the estimation or polling of nodal attributes when the network structure is not known but explorable. The nodal attribute was assumed to be binary, i.e., $x_i\in\{0,1\}$. For a set of randomly chosen nodes, $S$, the pollster may ask each node in $S$ about its own attribute for the estimation, which is intent polling. In contrast, the authors proposed neighborhood expectation polling (NEP) methods asking about the fraction of neighbors having $x_i=1$ among the node's neighbors. Then those fractions are averaged to give an estimation. Here, if nodes in $S$ are used, it is called a naive NEP method. Alternatively, each node in $S$ is used as the initial node of an $l$-step random walk along the network ($l\geq 1$), then the final node of the walk is asked to provide the fraction of its neighbors having $x_i=1$. Such an FP-based NEP method was proven to be more accurate than the other methods as it gives smaller mean-squared errors of the estimator when attributes of nodes are uncorrelated with each other. When high-degree nodes have one of two attributes more often, that attribute is overestimated. Experiments on five real-world social networks and on synthetic networks confirmed the tendency, most clearly when $|S|$ is small. The same authors~\cite{Nettasinghe2019Friendship} applied the FP to the estimation of the power-law exponent of the degree distribution too. Because sampling neighbors of random nodes draws more nodes from the tail than sampling random nodes, the FP lowers both the variance and mean-squared error of the maximum-likelihood estimate of the degree exponent, implying more accurate estimation. Later Nettasinghe et al.~\cite{Nettasinghe2024Estimating} devised an FP-based method for estimating exposure to information.

Alipourfard et al.~\cite{Alipourfard2020Friendship} extended the mentioned NEP methods to the directed social network derived from Twitter, where followers and followees are distinguished. They sampled random followees with probabilities proportional to their out-degrees to form a followee group, and the follower group was constructed by sampling random followers with probabilities proportional to their in-degrees. These sampled nodes are asked about how popular a hashtag is among their followees. Note that information flows from the followees to their followers, not the other way around. Since random followers have on average more followees than random nodes by the directed FP, the estimation using random followers shows the smaller variance of the popularity estimates of hashtags. In simulated polls on a subgraph of $5,409$ users, the mean-squared error of the NEP estimate was lower than that of intent polling for most of $500$ hashtags tested.

\subsection{Behavioral consequences of perception biases}\label{subsec:perception}

People tend to estimate the prevalence of an attribute in the population based on their immediate neighbors, which often induces biases. The resulting biases have long been studied in social psychology, e.g., in terms of false consensus~\cite{Mullen1985False, Marks1987Ten, Baer1991Biases} and pluralistic ignorance~\cite{Miller1987Pluralistic, Prentice1993Pluralistic}. These mechanisms attribute biases to individuals' own motivational or cognitive processes. In contrast, FPs (in particular, GFPs) could help us to focus on the structural cause of such perception biases or misperception. Since a node's neighbors are not a uniform sample, the node's observation of their neighborhood is already skewed before making any judgment on it. Such structural constraints can explain one's perception biases as well as their beliefs or behaviors to some extent. This may in turn impact beliefs or behaviors of their neighbors, their neighbors' neighbors and so on.

To the best of our knowledge, the first application of the GFP to understanding human behaviors was done by Pedersen~\cite{Pedersen1993Majority}, who scrutinized the majority fallacy. The majority fallacy indicates a phenomenon that one's own attribute is perceived as a majority in the population when it is not. It predicted that members in loosely integrated groups have a false impression of other members due to the lack of knowledge about them, which generates a false peer pressure, e.g., towards ``macho" roles among young males. The author surveyed $N=1,968$ students in and near Oslo, Norway in 1987 about their best friends and their own and their friends' drug uses. It was found that students having strong ties with friends tend to use more drugs, which could not be explained by the majority fallacy. In addition, the finding that students with more friends used more drug than average was essentially attributed to the GFP.

Lerman et al.~\cite{Lerman2016Majority} introduced the majority illusion, indicating that the attribute shared by a small fraction of high-degree nodes tends to be overrepresented so that the minority attribute appears like a majority to most nodes in the network. To quantify such a phenomenon, each node $i$ is assigned either $x_i=0$ (inactive) or $x_i=1$ (active). The probability that a node with degree $k$ has an active neighbor with $x'=1$ is written as
\begin{align}
    P(x'=1|k)=\sum_{k'}P(x'=1|k')P(k'|k),
\end{align}
where $P(x'=1|k')$ is the conditional probability of a node being active when its degree is $k'$. Then the fraction of nodes having more than half of their neighbors being active is calculated as
\begin{align}
    \mu =\sum_k P(k) \sum_{k/2<n\leq k}{k\choose n}P(x'=1|k)^n[1-P(x'=1|k)]^{k-n},
\end{align}
where ${k\choose n}$ is a binomial coefficient. That is, for a network of $N$ nodes, $\mu N$ nodes perceive that the majority of their neighbors is active. To demonstrate their results, the authors set the fraction of active nodes to be, e.g., $0.05$, and prepared scale-free and disassortative networks, e.g., with $P(k)\propto k^{-2.1}$ and $\rho_k=-0.35$. The value of $\mu$ could reach up to $0.6$--$0.8$ when degrees and attributes of nodes are positively correlated with each other, e.g., $\rho_{kx}\approx 0.4$. In other words, the minority opinion shared by only a small fraction of high-degree nodes can appear as a majority opinion to most, low-degree nodes. They tested their theory in six real-world networks too. 

Recently, the majority illusion has been translated into an algorithmic problem. Grandi et al.~\cite{Grandi2022Complexity} introduced the individual-level illusion by considering a binary attribute $x_i\in \{0,1\}$; a node $i$ is said to be under illusion if both the majority attribute among the $i$'s neighbors and the majority attribute in the entire network exist, while they are different from each other. They studied the computational complexity of checking, for a given $q\in [0,1]$, if a given network can be labeled with the binary attributes such that $qN$ nodes are under illusion. By explicitly considering tie cases when frequencies of two attributes are equal, Venema-Los et al.~\cite{Venema-Los2023Graph} investigated weak versions of majority illusion. Algorithmic methods of eliminating the majority illusion have also been explored either by adding and/or removing a fixed number of links~\cite{Grandi2022Complexity, Grandi2023Identifying, Dippel2025Eliminating} or by changing nodal attributes for a fixed number of nodes~\cite{Fioravantes2025Eliminating}.

Lee et al.~\cite{Lee2019Homophily} focused on the effects of the attribute homophily and minority group size on  perception biases in social networks. Let each node belong to either a minority group ($x_i=1$) or a majority group ($x_i=0$). The fraction of the minority in the network is denoted by $f_{\rm min}$ and it is less than $1/2$ by definition. The homophily is quantified by the probability $h$ that two neighboring nodes belong to the same group; $h>1/2$ means the homophily, while $h<1/2$ means the heterophily. The individual perception bias of a node $i$ is defined as the fraction of the minority nodes among $i$'s neighbors divided by $f_{\rm min}$:
\begin{align}
    b_i\equiv \frac{1}{f_{\rm min}}\frac{\sum_{j\in \Lambda_i} x_j}{k_i}.
\end{align}
Here $b_i=1$ indicates no bias\footnote{The individual perception bias $b_i$ can be seen as $x_{i,\rm nn}/\langle x\rangle$, which is similar to the friendship index in Eq.~\eqref{eq:FI_define} modified for the attribute but the denominator is the global average of $x$ instead of $x_i$.}. The group-level perception bias for the group $\alpha\in\{\rm min,maj\}$ is given as the average perception bias of nodes in the group:
\begin{align}
    B_{\alpha}\equiv \frac{1}{|G_\alpha|}\sum_{i\in G_\alpha}b_i,
\end{align}
where $G_\alpha$ is the set of nodes in the group $\alpha$. By the numerical simulation of the BA-homophily network model~\cite{Karimi2018Homophily}, generating a scale-free network with homophilic or heterophilic connections, they found that $B_{\rm min}$ ($B_{\rm maj}$) is an increasing (decreasing) function of $h$, and the intensity of the bias gets bigger with the smaller $f_{\rm min}$ for both $B_{\rm min}$ and $B_{\rm maj}$. That is, the more homophilic the network, the bigger perception biases occur for nodes in the minority, and the smaller perception biases occur for majority nodes. To test the validity of their theory, they surveyed participants in Germany ($N=99$), Republic of Korea ($N=100$), and the United States of America ($N=101$) on binary attributes such as smoking and attending worship. They found the same dependence of perception bias on $h$ and $f_{\rm min}$ in all three countries, suggesting that the homophilic structure of social networks and minority-group size could account for the biases held by people.

Jackson~\cite{Jackson2019Friendship} studied the role of the FP in the formation of social norms by analyzing three versions of a linear-quadratic game on a network: (i) In a society-wide game, a node $i$'s expected utility for its action, quantified by a positive value $x_i$, is assumed to be a function of its own $x_i$, preference for action $\theta_i$, and degree $k_i$ as well as of the expected action of the entire population, denoted by $\langle x\rangle$. The utility has a term of $x_ik_i\langle x\rangle$ with a positive coefficient, implying that a node's utility gets bigger when it has more neighbors and the population is more engaged in action. (ii) In a friend game, $\langle x\rangle$ in the utility function is replaced by the expected action of $i$'s neighbors only. (iii) In a naive game, nodes make decisions to maximize the expected utility in a friend game, while they experience the expected utility in a society-wide game. Assuming that preferences and degrees are independent, the author derived the analytical form of the action at equilibrium for each game to find that nodes act more in the friend game than in the society-wide game. In the friend game, the expected action of neighbors of nodes is higher than that of nodes themselves. These results are attributed to the FP. 

To empirically test if the FP does have behavioral consequences, Charness et al.~\cite{Charness2026Friendship} performed experiments where human subjects play a game of complements using a computer software in a laboratory. Each player $i$ is assigned to each node of a given network with $N=20$, and chooses an amount of action $x_i$ in each round of the game. The player is informed of how their payoff is determined by their own $x_i$ and degree $k_i$ as well as the average action of others, who are either the entire population except for $i$ (full information treatment) or $i$'s neighbors only (local information treatment). The payoff function is similar to the utility function in Ref.~\cite{Jackson2019Friendship}, but simplified. The authors found that players in the local information treatment chose higher actions than the full information counterparts, consistent with the expectation by the FP. During the game, players were relocated to other network positions, giving them opportunities to overcome the bias due to their positions; however, they failed to correct it.

Medhat and Iyer~\cite{Medhat2024Friendship, Medhat2024Exploring} devised a dynamical model to study how the FP-based social comparison can shape content-sharing or posting patterns in an online social network. In the model, each node $i$ of the network has a sharing rate $r_i(t)$ at time $t$ with an initial condition that $r_i(0)=1$. At each timestep, the node essentially compares the average sharing rate of its neighbors to the same average but weighted by degrees of its neighbors in such a way that higher-degree nodes have a stronger impact on their neighbors. In a simplest setup, when the ratio of the weighted average to the normal average exceeds some threshold value, the node stops sharing permanently; otherwise, the node keeps its initial sharing rate. Numerical simulations of the model show that the fraction of sharing nodes decreases to the equilibrium value.

Vilone and Polizzi~\cite{Vilone2024Modeling} studied an opinion dynamics model to understand the formation of misperception in networks. Each node $i$ has a private opinion or belief $x_i\in\{-1,1\}$ and the strength of the belief $\sigma_i\in (0,1]$, both of which are constant of time. The node also has an expressed opinion $y_i\in\{-1,0,1\}$, which can vary with time. As time goes by, a randomly chosen node expresses its private opinion with a probability $\sigma_i$, meaning that $y_i=x_i$, while with a probability $1-\sigma_i$, it expresses the majority opinion of its neighbors, i.e., $y_i$ takes the sign of $s_i\equiv \sum_{j\in\Lambda_i}y_j$ with a condition that $y_i=0$ if $s_i=0$. Two averages for expressed opinions are respectively defined as
\begin{align}
    &\langle y\rangle(t) \equiv \frac{1}{N}\sum_i y_i(t),\\
    &\langle y\rangle_{\rm ego}(t) \equiv \frac{1}{N}\sum_i \frac{1}{k_i}\sum_{j\in \Lambda_i}y_j(t).
\end{align}
The extent of misperception is quantified by the difference between the average belief $\langle x\rangle$ and $\langle y\rangle_{\rm ego}$ in the steady state. By the numerical simulations of the model on the scale-free network with $N=5,000$ and $P(k)\propto k^{-2.2}$, they found that the difference between $\langle y\rangle_{\rm ego}$ and $\langle y\rangle$ is overall smaller than the misperception defined above. It implies that the misperception can be attributed mainly to the nodes' own behaviors (endogenous) rather than to the network structure (exogenous). They also studied a variant of the model to emphasize on the role of high-degree but silent nodes (i.e., $y_i=0$) in helping the minor opinion to appear as a dominant one in the population.

Finally, Orlando et al.~\cite{Orlando2026Validating} devised a generative agent-based model by employing large language models (LLMs) to mimic human individuals in social media platforms. They confirmed that directed versions of FPs and GFPs emerged from the simulations using generative agents.

\section{Future works}\label{sec:discussion}

Despite the rapid developments on the topic of FPs in recent years, we find several promising research directions to deepen our understanding of FPs as well as to widen the scope of the field. Below we describe them one by one.

(i) Although various mathematical, computational, and statistical modeling approaches, including linear algebra and copula methods, have been successfully employed, more rigorous, analytical approaches to the FPs in networks are still required for the comprehensive understanding of FPs and their underlying mechanisms. 

(ii) Most empirical analyses have focused on the summary statistics of individual-level paradoxes, e.g., in terms of the network-level paradox holding probability $H$ in Eq.~\eqref{eq:FP_H_define}, or the network-level paradoxes, e.g., in terms of inequalities in Eqs.~\eqref{eq:FP_net_link_define} and~\eqref{eq:FP_net_ego_define}. In contrast, mesoscopic approaches, e.g., using $h(k)$ in Eq.~\eqref{eq:FP_hk_define}, are still largely unexplored. In particular, one may study if individual peer pressures in Eq.~\eqref{eq:FP_h_general} or individual neighborhood gaps in Eq.~\eqref{eq:deltak_i} are correlated between neighboring nodes, or how individual peer pressures or neighborhood gaps are distributed within or across communities in networks with community structure.
        
(iii) Novel network frameworks have emerged such as temporal networks~\cite{Holme2012Temporal, Holme2019Temporal, Masuda2020Guide}, multilayer networks~\cite{Kivela2014Multilayer, Boccaletti2014Structure, Aleta2019Multilayer, Artime2022Multilayer}, and higher-order networks~\cite{Lambiotte2019Networks, Battiston2020Networks, Bianconi2021Higherorder, Bick2023What}. However, little is known about the FPs on these kinds of networks. For example, one can ask the following questions: How would the FPs change over time in time-varying networks? How are the FPs in different layers of multilayer networks correlated with each other? Will hyperedges containing different numbers of nodes induce different degrees of FPs?

(iv) Dynamical processes taking place in networks, such as disease spreading~\cite{Barrat2008Dynamical, Pastor-Satorras2015Epidemic, Porter2016Dynamical, Masuda2017Temporal}, opinion formation~\cite{Castellano2009Statistical, Sirbu2017Opinion, Proskurnikov2017Tutorial, Starnini2026Opinion, Caldarelli2026Physics}, and evolution of cooperation~\cite{Perc2013Evolutionary, Takacs2021Networks}, could be affected by the FPs. For example, Lee et al.~\cite{Lee2019Impact} studied the FP-based opinion dynamics in networks, and Santos et al.~\cite{Santos2021Biased} demonstrated how perception biases could lead to the suboptimal collective action in nonlinear public goods games. On the other hand, the FP-based perceptions could be influenced by or even coevolving with such dynamical processes. Further, one can study the case when the network topology, the FP-based individual perception, and the dynamical process coevolve together. Note that researchers have studied epidemic spreading in a population, where individuals are aware of the disease and can change their behaviors accordingly~\cite{Jo2006Immunization, Funk2009Spread, Granell2013Dynamical, DaSilva2019Epidemic}, as well as coevolutionary dynamics of a network and individual nodes' perception of it~\cite{Jo2015Coevolution, Jo2016Dynamical}.
    
(v) It is also expected that the graph representation or network embedding in machine learning~\cite{Hamilton2017Representation, Chen2020Graph, Khoshraftar2024Survey} can benefit from the FPs. For example, Kojaku et al.~\cite{Kojaku2021Residual2Vec} paid attention to the sampling bias due to the FP and introduced debiasing of the random walk on the network for graph representation learning.

\begin{acknowledgments}
The authors thank Young-Ho Eom for bringing the friendship paradox to their attention and for the longstanding collaboration. H.-H.J. gratefully acknowledges Naoki Masuda for the invitation to visit the University of Michigan, where part of this work was conducted, as well as financial support by the National Research Foundation of Korea (NRF) grant funded by the Korea government (MSIT) (RS-2026-25476645).

The authors used Claude Opus 5 to edit the English of the initial draft; however, the final manuscript was fully verified by them. 
\end{acknowledgments}

%

\end{document}